\documentclass[sigconf,nonacm]{acmart}
\setcopyright{none}
\usepackage{xurl}  \usepackage{graphics}
\usepackage{amsmath}	   \usepackage{shadow}        \usepackage{hhline}        \usepackage{tabularx}      \usepackage{latexsym}      \usepackage{fancyhdr}      \usepackage{wrapfig}	   \usepackage{capt-of}	   \usepackage{verbatim}
\usepackage{xspace}		   \usepackage{balance}     
\usepackage{algpseudocode} \usepackage{algorithm}
\usepackage{outlines}      \usepackage[gen]{eurosym}  \usepackage{subcaption}    \usepackage{color}
\usepackage{enumitem}
\usepackage{multirow}
\usepackage{makecell}

\usepackage{xfp}

\usepackage{shortcuts}

\renewcommand{\paragraph}{\vspace{1pt}\noindent\textbf}
\begin{document}

\setlength{\parskip}{3.5pt}

\title[Identifying Cybercrime Financial Relationships in Bitcoin through Back-and-Forth Exploration]{Watch Your Back: Identifying Cybercrime Financial Relationships in Bitcoin through Back-and-Forth Exploration}

\author{Gibran Gomez}
\affiliation{  \institution{IMDEA Software Institute}
  \institution{Universidad Polit\'{e}cnica de Madrid}
  \country{}
}
\email{gibran.gomez@imdea.org}

\author{Pedro Moreno-Sanchez}
\affiliation{  \institution{IMDEA Software Institute}
  \country{Madrid, Spain}
}
\email{pedro.moreno@imdea.org}

\author{Juan Caballero}
\affiliation{  \institution{IMDEA Software Institute}
  \country{Madrid, Spain}
}
\email{juan.caballero@imdea.org}

\pagenumbering{arabic}

\begin{abstract}

Cybercriminals often leverage Bitcoin for their illicit activities.
In this work, we propose \exploration, a novel 
automated Bitcoin transaction tracing technique to identify
cybercrime financial relationships. 
Given seed addresses belonging to a cybercrime campaign,
it outputs a \graph,
and identifies paths corresponding to relationships 
between the campaign under study and 
external services and other cybercrime campaigns.
\Exploration provides two key contributions.
First, it explores both forward and backwards, 
instead of only forward as done by prior work, 
enabling the discovery of relationships that cannot be found by 
only exploring forward (e.g., deposits from clients of a mixer).
Second, it prevents graph explosion by combining
a tagging database with a machine learning
classifier for identifying addresses belonging to exchanges. 

We evaluate \exploration on \numoperations malware families.
We build oracles for 4 families using Bitcoin for C\&C and use them
to demonstrate that \exploration identifies 13 C\&C signaling addresses
missed by prior work, 8 of which are fundamentally missed by 
forward-only explorations. 
Our approach uncovers a wealth of 
services used by the malware
including 44 exchanges, 11 gambling sites, 5 payment service providers,
4 underground markets, 4 mining pools, and 2 mixers.
In 4 families, the relations include new attribution points 
missed by forward-only explorations.
It also identifies relationships between the malware families and other 
cybercrime campaigns, highlighting how some malware operators 
participate in a variety of cybercriminal activities.

\end{abstract}

\ccsdesc[500]{Security and privacy-Malware and its mitigation}

\keywords{Cybercrime Financial Relations, Blockchain, Malware, Clipper}

\maketitle

\section{Introduction}
\label{sec:intro}

Cryptocurrencies such as Bitcoin are attractive for cybercriminals due to 
their decentralized nature, 
(pseudo-)anonymity,
irreversible transactions, and 
the ease to buy and sell them.
However, some cryptocurrencies such as Bitcoin 
have a public transaction ledger,
which has enabled the analysis of 
diverse cybercrime activities such as 
ransomware~\cite{cerber,behindLiao,economicConti,trackingHuang,ransomwareClouston,bitiodineSpagnoulo},
thefts~\cite{fistfulMeiklejohn},
scams~\cite{spamsPaquetClouston,ponziBartoletti,fistfulMeiklejohn},
human trafficking~\cite{backpagePortnoff},
hidden marketplaces~\cite{travelingChristin,cybercriminalLee,dreadRon},
money laundering~\cite{inquiryMoser}, and
cryptojacking~\cite{sokCryptojacking}.

A key component of a Bitcoin cybercrime investigation is the analysis of 
how money flows, e.g., how cybercriminals finance their activities and 
where the profits of those activities go.
Identifying the financial relationships
(relations for short) between a cybercrime campaign and 
external services (e.g., exchanges, mixers, underground markets), 
or with other cybercriminal campaigns, is fundamental in attribution. 
For example, if a coin flow is identified between a malicious campaign 
of interest and a service (e.g., an exchange) 
with know-your-customer (KYC) requirements~\cite{gill2004preventing}, 
the service can be leveraged as an attribution point 
by law enforcement to identify the perpetrators. 
Moreover, identifying relations between cybercrime campaigns, which 
were initially considered independent, sheds light into services 
supporting cybercriminal activities~\cite{clayton2015concentrating},
and their operators, which become then targets for takedown ~\cite{btceTakedown,bestmixerTakedown,bitcoinfogTakedown}.
Yet another application is identifying the malicious clients of a 
specific service, 
e.g., a mixer suspected of laundering money from illicit activities.
This information has been fundamental in past takedown actions, 
by justifying to a judge that the service is a hotspot of malicious activity. 
For example, the arrest warrant for the \emph{Bitcoin Fog} mixer 
operator was justified on its abuse for money laundering by 
multiple underground markets~\cite{bitcoinfogAffidavit}.

A common technique for analyzing Bitcoin abuse 
is \emph{\tracing},
i.e., tracing the flow of BTCs starting from a given set of \textit{seed addresses} known to belong to a cybercrime campaign 
~\cite{trackingHuang,ransomwareClouston,spamsPaquetClouston,cybercriminalLee}.
Examples of seed addresses are those in ransom notes 
(e.g.,~\cite{trackingHuang,ransomwareClouston}) and
those in scam emails
(e.g.,~\cite{spamsPaquetClouston}).
But, \tracing has two fundamental limitations.
First, it only considers forward propagation of the funds from the seeds, 
i.e., it does not explore backwards.
For example, if the seed is used as input to a transaction with two inputs 
and one output, it traces the output, but it does not trace 
(backwards) the other input, which we can also attribute to the 
campaign since multiple inputs of the same transaction are associated to the same 
actor (i.e., multiple-input heuristic)~\cite{fistfulMeiklejohn}.
Second, if the output address receives another deposit transaction
(where the seed is not an input) that deposit transaction is ignored.  
Thus, forward-only exploration
may often miss fundamental relations.
For instance, malware that uses the Bitcoin blockchain
for signaling C\&C servers (e.g.,~\cite{cerber,pony})
needs to seed the signaling addresses with funds.
Failure to explore backwards from the signaling addresses 
may miss the exchange used to seed the C\&C signaling that 
can be leveraged as an attribution point.
Furthermore, since the funds are consumed to pay the fees of 
the signaling transactions, there may not be a cash out that 
forward-only exploration can identify.
Another situation is when examining the clients of a 
potentially malicious service, e.g., a mixer. 
Forward-only exploration from the mixer addresses is useless as 
the mixer's goal is precisely to prevent forward tracing. 
However, backwards exploration from the mixer's addresses allows 
identifying the clients that leverage the mixer, 
which can be used to justify an arrest warrant~\cite{bitcoinfogAffidavit}.
Last but not least, existing approaches perform shallow explorations to prevent 
the transaction graph from exploding in size.
We posit that the key challenge to prevent such explosion, 
and also for identifying relations, 
is to identify \emph{change of ownership},
i.e., when BTCs change hands from the operators of the campaign under study
to other entities.
Prior works detect change of ownership using commercial databases of 
tagged addresses~\cite{chainalysis,elliptic}.
However, such databases have limited coverage, 
which can make the graph explode.
For example, failing to identify that the cybercriminals cashed out 
at an exchange would include a massive amount of transactions 
from other innocent clients of the exchange into the transaction graph. 

To address these limitations, we contribute the following: 

\begin{enumerate}[label=\textbullet,wide=\parindent]
	\item We propose \emph{\exploration}, 
a novel automated transaction tracing technique to identify 
cybercrime financial relationships. Given a set of seed 
addresses belonging to a cybercrime campaign, 
it outputs a \graph 
and it extracts paths in the \graph that capture relations to other 
entities and campaigns.
The \graph can be used as evidence of the relations identified,
and can be shared with law enforcement and other analysts
for independent validation,
or with a judge to justify a warrant.
Moreover, the relations could be publicly shared, preventing services 
(e.g., exchanges) that interact with the cybercrime campaigns 
from claiming that they were unaware of the abuse. 
\Exploration provides two key contributions over current \tracing 
techniques. 
First, it explores both forward and backwards, 
enabling the discovery of important relations missed otherwise.
Second, it combines the use of a tagging database with a machine learning 
classifier to identify addresses belonging to exchanges.
The classifier can detect previously unknown (i.e., untagged) 
exchange addresses, preventing graph explosion.

	\item We propose \emph{\oracles}, 
binary classifiers that can identify specific types of malicious addresses 
using sequences of distinctive transactions. 
We build \oracles for four malware families that use the Bitcoin
blockchain for signaling C\&C servers. We use them to identify signaling addresses in the \graphs, 
demonstrating that \exploration can discover 13 signaling addresses 
missed by previous works~\cite{cerber,pony, gluptebaTakedown} and that
backwards exploration is key for their discovery. 

\item We build a database of \ntags tagged addresses from open sources
and expand it into \nctags tagged addresses through multi-input clustering.
We show that tag databases need to handle \emph{double ownership},
i.e., Bitcoin addresses with more than one owner.
Double ownership is common in services offering online wallets,
which create an address for a specific user.
In that scenario the service owns the address
(i.e., has the private key),
but the address is handled (and thus indirectly owned)
by the user for which it is created.

\item We apply \exploration on \numoperations malware families:
\numoperationsclipper clippers 
that replace addresses in the OS clipboard
of the infected host with an address controlled by the malware operators, 
\numoperationsransomware ransomware, and 
\numoperationsmalware families using Bitcoin for C\&C.
\Exploration finds relations in 93\% of the explorations. For 13\% of the families, it identifies new attribution points, 
missed by forward-only explorations.
It uncovers that 23 replacement addresses for 9 clippers 
are online wallets in exchanges (double ownership).
This shows that clippers, a largely overlooked, but highly effective, 
malware class, often directly cash out the stolen funds without intermediate
steps. 
It is surprising that the abusive nature of those addresses 
goes unnoticed for the exchanges.
Overall, the most common relations are with 44 exchanges.
Exchanges used by multiple families 
(e.g. \tag{localbitcoins}, \tag{coinpayments}, \tag{coinbase},
\tag{poloniex}, \tag{cryptonator}, \tag{btc-e})
are often not the largest ones, 
which could indicate a more lax approach to abuse detection. 
In fact, one of those exchanges has already been taken down~\cite{btceTakedown}.
Other identified relations are 
2 mixers (\tag{wasabi}, \tag{bitcoinfog}), 
the latter being recently taken down 
for money laundering~\cite{bitcoinfogTakedown};
11 gambling sites likely also used for money laundering;
5 payment services used for acquiring resources;
4 mining pools, and
4 underground markets.
In 7 explorations, relations are found between a
malware family and other cybercrime entities.
We observe that some malware operators participate in a variety of 
cybercriminal activities. 
For example, the \tag{MrPr0gr4mmer} operator sells spam tools, 
sends sextortion scam campaigns, and handles a clipper.
We also identify clippers run by the same operators
and 
observe malware operators using profits to upgrade their software 
arsenal (e.g., buying licenses for malware kits and spam tools) 
and for seeding new C\&C signaling approaches.
Finally, we perform a backwards-only exploration on the \tag{bitcoinfog} mixer 
identifying 15 underground markets, 130 other services, and 
15 malicious operations that leveraged it as clients.

\item We release our exploration code and tag database~\cite{watchyourback}.

\end{enumerate}

\section{State of the Art}
\label{sec:overview}

\subsection{Bitcoin Address Ownership}
\label{sec:ownership}

The notion of \emph{ownership} of a Bitcoin address is associated 
to the user who possesses the corresponding signing key  
and can thus authorize a transfer of coins from that address. 
A user may own multiple addresses. 
For example, a (privacy-aware) user may create a new
address for every payment he receives.

\paragraph{Multi-input clustering.}
We leverage a popular technique to identify Bitcoin addresses 
belonging to the same owner called multi-input (MI) clustering~\cite{bitcoin,evaluatingAndroulaki,quantitativeRon,fistfulMeiklejohn}.
It assumes that input addresses to the same transaction
have the same owner because their private keys are 
used together to sign the transaction.
Clusters can be created transitively. 
If a transaction has addresses $A$ and $B$ as inputs, and 
another transaction has $B$ and $C$ as inputs, then $A$, $B$ and $C$ 
are all clustered together and have the same owner. 
One exception 
are CoinJoin~\cite{coinjoinMaxwell} transactions
where a group of users create a single transaction 
that simultaneously spends all their inputs into a (shuffled) list of 
outputs. 
For this reason, before computing MI clustering, 
we apply proposed heuristics to identify 
CoinJoin transactions~\cite{cookieGoldfeder,blocksci}.
There exist other address clustering heuristics
such as shadow address~\cite{evaluatingAndroulaki},
change address~\cite{fistfulMeiklejohn},
and its variants (peeling chain, power of ten, address reuse)~\cite{blocksci}.
However, those techniques rely on usage patterns that change over time and 
can lead to high false positives.

\paragraph{Tagging.} 
While MI clustering identifies addresses belonging to the 
same owner, it does not determine who the owner is. 
For this, commercial services 
(e.g.,~\cite{chainalysis,elliptic}) 
provide paid access to databases of addresses tagged 
with a label stating their owner or usage. 
Instead, we leverage public resources to build our own tag database.
An intrinsic limitation of tag databases is that
they only include a small fraction of Bitcoin addresses, 
having thus a low coverage. 
In this work, we show that when tagging addresses, 
it is fundamental to handle \emph{double ownership}, 
i.e., that a Bitcoin address may have more than one owner. 
In the double ownership scheme, a first user 
(i.e., the \emph{owner}) 
possesses the signing key corresponding to the address and accepts 
transaction requests (possibly authenticated with e.g., username and password) 
from a second user 
(i.e., the \emph{beneficiary}) 
who is thereby the one governing where the coins are transferred and when. 
Note that the owner does not share the signing key to the 
beneficiary, so the beneficiary cannot produce transactions on its own. 
This double ownership scheme 
is used for instance by some exchanges and online wallets. 
These services, on user's request to deposit its coins, 
may create a fresh address (and the corresponding signing key) 
where afterwards the user's coins are deposited. 
Handling double ownership is fundamental when collecting tagged addresses 
from multiple public sources. 
For example, the same address may be tagged in one 
source as belonging to an exchange and in another 
source as being the replacement address for a clipper malware. 
Without taking double ownership into account that would create a conflict 
in assigning ownership to the address. 
Instead, the address should be tagged as belonging to an exchange 
(i.e., the owner) that has assigned that address to an exchange user 
(i.e., the beneficiary) who happens to be the clipper's operator.
Our tag database construction is detailed in Section~\ref{sec:tagging}.

\paragraph{Exchange address classifier.}
We propose a machine learning classifier to identify 
addresses belonging to exchanges, 
even if the specific exchange it belongs to was never tagged and is 
therefore not part of the training dataset. 
Our exchange classifier prevents graph explosion due to limited tagging 
coverage, 
but can only determine a given address belongs to an 
exchange, not to which specific exchange.
Our exchange classifier is detailed in Section~\ref{sec:classifier}.

\paragraph{Operation oracles.}
Previous works build functions that given a ransomware collection address
distinguish victim payments by examining if deposit amounts are within 
payment ranges used by the 
family~\cite{behindLiao,economicConti,ransomwareClouston,trackingHuang}. 
We expand this concept by proposing \emph{operation oracles}, 
binary classifiers that identify specific types of malicious addresses 
(performing distinctive transactions)
by applying rules to the sequence of transactions involving the address. 
We build operation oracles, 
detailed in Section~\ref{sec:evaluation},
for three malware families  
that leverage the Bitcoin blockchain to signal C\&C endpoints.
Our operation oracles consider transaction sequences, 
e.g., that address A sends a payment to address B and soon after address 
B returns the same amount, minus the transaction fee, back to A.
Considering multiple transactions and dependencies between them 
makes the classifier more distinctive than 
considering only transaction values, reducing false positives (FPs).

\subsection{Transaction Tracing}
\label{sec:tracing}

\Tracing focuses on the outgoing transactions performed 
by the seeds and transitively by their destination addresses 
\cite{trackingHuang,ransomwareClouston,spamsPaquetClouston,cybercriminalLee}.
A main limitation of \tracing is that it only explores forward, 
but it does not explore backwards. 
In particular, in \tracing, if an address is being traced, 
then the outputs of all transactions where that address is used as 
input will also be traced. 
But, other inputs in those transactions will not be traced. 
In addition, deposit transactions into a traced address,
where the traced address appears as output, 
but no input is currently traced, 
are not traced either. 
Due to this limitation, 
\tracing cannot discover relations only reachable 
by exploring backwards. 
For example, \tracing starting from addresses belonging to a mixer 
makes the exploration explode due to the mixing behavior, 
while exploring backwards can reveal the clients using the mixer.
In another example, 
exploring backwards from a C\&C signaling address can reveal 
the exchange used to fund the signaling. 
But, \tracing may not reveal such attribution 
point as the funds are consumed in the signaling.

To address this limitation we propose a novel \emph{\exploration}, 
which traces forward and backward. 
In a bit more detail, 
beginning from a seed, our \exploration retrieves
deposit addresses of incoming transactions,
withdrawal addresses of incoming transactions,
deposit addresses of outgoing transactions, and 
withdrawal addresses of outgoing transactions.

\begin{figure}
  \centering
  \includegraphics[width=\columnwidth]{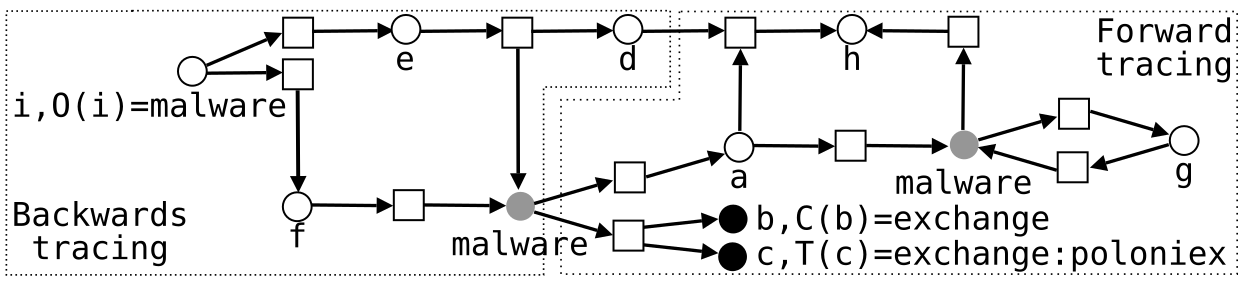}
  \caption{Example transaction graph from two malicious seeds in gray.
\Tracing would discover the right subgraph, 
while \exploration would discover the whole graph including a new 
malicious address.}
  \label{fig:exp}
\end{figure}

We illustrate the differences between \tracing and \exploration using 
the example transaction graph in Figure~\ref{fig:exp}.
In the graph, circle nodes correspond to addresses and square nodes 
to transactions. Nodes filled in black are those belonging to a cluster tagged as exchange.
Edges capture coins flowing from an input address to a transaction, 
or from a transaction to an output address.
The exploration starts from the two seeds in gray. 
The graph is split in two.  
The right side corresponds to the graph that \tracing would produce 
and the left side is the additional subgraph that \exploration 
would additionally discover, including a new malware address.
There are two points were both explorations differ. 
First, \exploration traces deposit addresses to the leftmost seed, 
including addresses $f$, $e$, and $d$ into the graph.
Second, once address $a$ has been discovered through a forward step from 
the leftmost seed, the transaction where $a$ is an input and $h$ is an 
output will be traced. 
At that point, \exploration would include address $d$ in the tracing
while \tracing would not.

One risk of backwards exploration is that any entity can deposit  
BTCs into any address. 
If a benign entity sends funds to a malicious address, 
the backwards exploration may consider the sender address part of the malicious 
campaign.
Since addresses in ransom notes and scam emails receive victim's payments and 
are typically seeds for our \exploration, 
\tool has an option to prevent exploration backwards from the seeds. 
Note that even if this option is used, \exploration 
still explores backwards at intermediate nodes.
For example, in Figure~\ref{fig:exp}, even if backwards exploration on the 
seeds is disabled, the final graph would be the same because the transaction 
between $a$ and $h$ would bring $d$ into the graph, and successive 
backwards steps would then include $e$, $i$, and $f$.
There are two other cases where backwards exploration can be problematic. 
One case are \emph{forced address reuse} (or \emph{dust}) attacks
where an entity performs micro-payments to addresses it does not own, 
hoping to incentivize the receivers to move those 
small amounts in later transactions~\cite{trackingHuang}.
In this work, we use some filtering heuristics to identify and remove 
forced address reuse transactions.
The other case is researchers trying to hijack the C\&C 
infrastructure of a malware by sending payments to the signaling 
addresses~\cite{pony}. 
In such cases, researchers are likely to specify their transactions 
to avoid being incorrectly linked to the malicious 
campaign, allowing their filtering.

\paragraph{Change of ownership.}
A fundamental challenge when tracing a malicious campaign 
is to detect change of ownership, 
i.e., addresses that belong to different entities. 
A critical change of ownership is when BTCs flow into exchanges, 
e.g., cybercriminals cashing out their profits.
Failure to detect an exchange address would make the graph explode by 
introducing thousands (or even millions) of transactions from other 
clients of the exchange.
Detecting change of ownership is also fundamental to analyze relationships 
between the seed owners and external entities.
In this case, failure to detect an exchange address would make it 
look like there exist relations between unrelated cybercriminal campaigns 
that simply use the same exchange.
For example, in Figure~\ref{fig:exp} address $b$ is identified as 
an exchange by our ML classifier, while $c$ is identified as 
belonging to the \tag{poloniex} exchange through a previously available tag. 
Failure to detect these addresses would have made the graph explode.

\begin{figure}[t]
  \centering
  \includegraphics[width=\columnwidth]{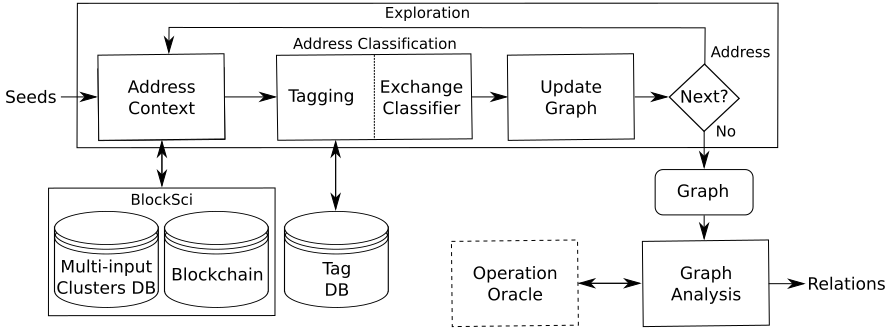}
  \caption{\Exploration architecture.}
  \label{fig:arch}
\end{figure}

\section{\EXploration}
\label{sec:exploration}

Given seed Bitcoin addresses belonging to a cybercrime campaign, 
\exploration outputs a \graph and identifies paths in the graph that 
capture relations to other entities and campaigns.
As illustrated in Figure~\ref{fig:exp}, 
nodes in the \graph are of two types: addresses and transactions. 
Edges are directed; 
they may connect an output slot in a transaction to the address where funds 
are received, or connect an address to the input slot in a 
transaction where funds are sent to.
If the same address appears several times as input or output of the same 
transaction, a single edge connects address and transaction.
Address nodes are reused if the same address appears in multiple transactions.
Address nodes have attributes capturing 
their alphanumeric address string,
the list of tags associated to the address (if any), 
the identifier of the MI cluster the address belongs to, and the address type 
(e.g., if detected by the exchange classifier or by the \oracle).
Transaction nodes have as attributes the transaction identifier and 
the total BTC amount transferred.
Edges have as attributes the transaction ID, the slot number, and 
the BTC amount transferred in that slot.

The \exploration architecture is shown in Figure~\ref{fig:arch}.
It performs an iterative process.
At each iteration, it selects a Bitcoin address from the worklist and 
performs three steps. 
First, it obtains the address context by 
identifying the MI-cluster the address belongs to and 
extracting from the blockchain its deposit and withdrawal transactions. 
The blockchain data is processed using the BlockSci framework~\cite{blocksci}.
Before launching any exploration, BlockSci is used to 
parse the blockchain data up to the desired block height and 
to pre-compute the MI clustering results. 
Second, it classifies the address to check for change of ownership 
using the input tag database (detailed in Section~\ref{sec:tagging}) 
and the exchange classifier (detailed in Section~\ref{sec:classifier}). 
Finally, it updates the \graph with the obtained transactions and addresses; 
and adds to the worklist any new addresses that need to be explored.
The process starts by placing the seeds in the worklist and
finishes when the worklist is empty, or 
a configurable exploration limit is reached.
At the end of this iterative process the \graph is output.

The graph analysis module takes as input the \graph and 
an optional \oracle. It outputs relations between the campaign under study 
and other entities in the graph. 
First, it tags addresses satisfying the oracle, 
e.g., as a C\&C signaling address.
Then it identifies relations, 
which are paths between a seed and a tagged address that 
satisfy the following constraints.
First, the found tag should differ from the seeds' tag.
Second, the path should be directed from the seed to the tagged address, or
from the tagged address to the seed, 
i.e., it should not mix edges with opposite directions.
This constraint makes sure that the relation captures money flow
from the campaign under analysis to an external entity or vice-versa.
Third, addresses identified by the exchange classifier,
for which no tags were available, are excluded since it is not 
known to what specific exchanges they belong to.

\paragraph{Address context.}
At each iteration, the address being explored is used to query BlockSci 
to obtain the ID of the MI cluster to which the address belongs.
If the cluster has already been examined, 
e.g., if detected as an exchange when examining a previous address in the 
cluster, 
its classification results are recovered.
Then, BlockSci is queried again to obtain the list of all 
transactions involving the address. 
CoinJoin and forced address reuse transactions are filtered from the exploration. 
CoinJoin transactions are identified using a set of 
previously proposed heuristics~\cite{cookieGoldfeder,blocksci} 
implemented by BlockSci.
To identify force address reuse transactions we use our own heuristic, 
which looks for transactions that send the same amount to a large number 
of (at least 100) output addresses. 
Next, it obtains from BlockSci the list of 
addresses involved in the remaining transactions, 
both in input and output slots.
Transactions and addresses are added to the \graph 
and passed to the address classification that determines
which addresses should be explored in later iterations.

\paragraph{Address classification.}
Each address received from context extraction
is examined for change of ownership. 
Addresses identified as belonging to other entities,
either through tags or by the exchange classifier, 
will be labeled as such and will not be traced.
The rest will be added to the worklist to be traced.
First, the address is queried to the tag database.
If tagged, and if its tag category identifies it as a 
service (e.g., exchange, onlinewallet, mixer, tormarket, gambling), 
the address is not added to the worklist.
There is no need to keep exploring the address since that would explore 
the service, 
introducing many extraneous nodes in the \graph.
Second, the ML exchange classifier, detailed in Section~\ref{sec:classifier}, 
is applied to the address.
If detected, the address is not added to the worklist. 
As an optimization, the ID of the address' MI-cluster is cached, so that 
any other address in the same cluster
can be directly marked as an exchange without going through the 
classification.
Addresses added to the worklist are 
assigned a priority inversely proportional 
to the total number of input and output slots in their transactions.
The address with highest priority is selected and another 
iteration starts.

The intuition behind the priority is that 
addresses with many transactions, or with very large transactions,
could belong to services not in the tag database or to 
false negatives of the exchange classifier. 
To avoid the exploration being stuck exploring those addresses, 
their priority is lowered so that they are explored at the end.
The priority only affects the order of the exploration
as long as the exploration runs to its end. 
However, if the user configures a maximum exploration limit, 
high volume addresses may not be explored before the limit 
is reached, which could make some relations to be missed.

\begin{table}
\centering
\scriptsize
\begin{tabular}{|l|l|rr|rr|}
\cline{3-6}
\multicolumn{2}{c|}{} & \multicolumn{2}{c|}{\textbf{Addresses}} & \multicolumn{2}{c|}{\textbf{Multi-Input Clustering}} \\
\hline
\textbf{Category} &	\textbf{Class} & \textbf{All} & \textbf{wTXES} & \textbf{Clusters} & \textbf{Addresses} \\
\hline

sextortion & Abuse & 62,600 & 407 & 171 & 1,131 \\
miscabuse & Abuse & 57,031 & 12,018 & 7,568 & 8,656,754 \\
mining & Service & 37,454 & 37,453 & 29,526 & 594,609 \\
mixuser & Individ. & 36,548 & 36,548 & 7,308 & 36,860 \\
ransomware & Malware & 17,668 & 17,447 & 10,348 & 21,160 \\
clipper & Malware & 9,513 & 464 & 105 & 1,376 \\
username & Individ. & 6,618 & 5,399 & 5,110 & 153,826 \\
exchange & Service & 1,301 & 1,299 & 737 & 75,257,530 \\
terrorism & Abuse & 275 & 275 & 48 & 306 \\
theft & Abuse & 221 & 220 & 129 & 2,072,269 \\
gambling & Service & 182 & 182 & 106 & 8,800,225 \\
onlinewallet & Service & 140 & 129 & 37 & 4,612,732 \\
defi & Service & 93 & 17 & 17 & 15,945 \\
scam & Abuse & 72 & 61 & 32 & 314,537 \\
mixer & Service & 56 & 55 & 52 & 2,137,493 \\
ponzi & Abuse & 54 & 54 & 41 & 20,908 \\
service & Service & 50 & 50 & 43 & 1,293,660 \\
malware & Malware & 19 & 18 & 17 & 38 \\
tormarket & Service & 19 & 19 & 19 & 2,327,272 \\
payment & Service & 15 & 15 & 12 & 9,717,967 \\
donation & Benign & 14 & 12 & 8 & 217 \\
state-sponsored & Abuse & 14 & 14 & 8 & 66,855 \\
drugtrafficking & Abuse & 11 & 11 & 2 & 38 \\
bankingtrojan & Malware & 3 & 1 & 1 & 1 \\
cryptojacking & Malware & 3 & 3 & 2 & 25 \\
miscbenign & Benign & 3 & 3 & 1 & 2 \\
webskimming & Malware & 3 & 3 & 3 & 6 \\
usgov & Benign & 2 & 2 & 2 & 3 \\
\hline
\multicolumn{1}{c|}{} & \multicolumn{1}{l|}{All} & 229,982 & 112,179 & 61,453 & 116,103,745 \\

\cline{2-6}

\ignore{
	sextortion & 62,539 & 348 & 122 & 1,753,120 \\
	miscabuse & 57,302 & 12,117 & 7816 & 107,576,643 \\
	mining & 37,453 & 37452 & 29529 & 15,820,506 \\
	mixuser & 36,548 & 36548 & 7308 & 36,860 \\
	ransomware & 17,627 & 17447 & 10352 & 14,347,286 \\
	clipper & 9,453 & 414 & 83 & 28,117,255 \\
	username & 6,614 & 5395 & 5112 & 36,573,610 \\
	exchange & 865 & 865 & 515 & 32,844,163 \\
	theft & 242 & 242 & 151 & 25,756,316 \\
	gambling & 158 & 158 & 97 & 3,286,539 \\
	ponzi & 53 & 53 & 46 & 14,503,196 \\
	onlinewallet & 52 & 52 & 20 & 2,011,429 \\
	mixer & 46 & 46 & 44 & 2,128,745 \\
	service & 40 & 40 & 38 & 1,061,722 \\
	scam & 39 & 38 & 23 & 1,252,965 \\
	tormarket & 16 & 16 & 16 & 2,327,192 \\
	donation & 9 & 9 & 8 & 217 \\
	malware & 8 & 8 & 8 & 25 \\
	miscbenign & 2 & 2 & 2 & 13,171,445 \\
	cryptojacking & 1 & 1 & 1 & 2 \\
	usgov & 1 & 1 & 1 & 1 \\
\hline
	Total & 230,450 & 111,461 & 61,292 & - \\
\hline
}
\end{tabular}
\caption{Tag database summary.}
\label{table:tags}
\vspace{-0.8cm}
\end{table}

\section{Tagging}
\label{sec:tagging}

Tag database construction comprises of two steps.
First, we collect \ntags addresses from public sources and
tag them using their context.
Then, we propagate the address tags to their MI clusters 
obtaining \nctags tagged addresses.
After each step there is a disambiguation process to resolve any 
conflicts due to multiple tags assigned to the same address or cluster.
We explore a variety of publicly available sources to 
tag Bitcoin addresses including 
databases of already tagged addresses, 
(e.g., WalletExplorer tags~\cite{walletexplorer} 
available in GraphSense~\cite{graphsenseHaslhofer}),  
web pages where services disclose their addresses 
(e.g., the SatoshiDice gambling service~\cite{satoshiDice}),
forums where users rate each other
(e.g., BitcoinTalk~\cite{bitcoinTalk}, Bitcoin-OTC~\cite{bitcoin-otc}),
crowd-sourced databases where users report abusive addresses 
(e.g., Bitcoin Abuse~\cite{bitcoinAbuse}), and security blog articles~\cite{goodfatr}. 
The different sources have little overlap and each source provides
some addresses that do not appear in any other source.
Thus, they all contribute to making the tag database as complete as possible.
Some of these sources such as WalletExplorer and Bitcoin-OTC
provide their own tagging database,
which can be automatically imported into our platform.
For others like BitcoinTalk, tags are produced by manually inspecting 
specific reports. 
For Bitcoin Abuse, the category is provided, but selecting the label requires 
manually inspecting the reports. 
Updating the tagging database needs to be done periodically and is the 
only step in our approach that requires human involvement.
In future work, we plan to explore automating this step using NLP.
For the interested reader, Table~\ref{tab:tags_source} in the Appendix details
the top 20 source URLs by number of tagged addresses.

Each database entry is a 6-tuple 
(\emph{address}, \emph{class}, \emph{category}, \emph{label}, 
\emph{subtype}, \emph{urls}). 
There are 28 categories, detailed in Table~\ref{table:tags},
which can be grouped into 6 classes:
malware, abuse, services, individuals, and benign. Label captures the ownership, 
i.e., the specific malware family, service, or individual 
that owns the address.
Subtype is an optional component that details the usage of the address,
e.g., if it is the hot wallet of an exchange or 
a malware C\&C signaling address.
The final component is the list of source URLs that discuss the address.
For example, address \emph{1NDyJt} has a category 
\emph{exchange}, label \emph{binance} (the specific exchange), 
hot wallet as subtype, 
and a URL to the WalletExplorer webpage that describes it.
Throughout the paper, when we refer to a tag 
we mean its category and label, 
e.g., \emph{exchange:binance}. 
We identify addresses by their first 6 characters,
but provide full addresses in Table~\ref{tab:mapping} in the Appendix.

We initially assign a tag to an address for each source that mentions it. 
Then, we select a unique address tag by applying the following criteria for
resolving potential ambiguity in addresses with multiple tags.
If an address has multiple tags with the same category and label, 
we favor the tag that provides more information, 
i.e., the one with a subtype, or the longest one otherwise.
If an address has multiple tags with different category and label pairs, 
we favor tags not coming from Bitcoin Abuse because 
categories in that service are assigned by users with varying expertise and 
it is not rare that they are wrong.
If an address has multiple tags of the mining category, 
we merge its labels to capture all the mining pools where the miner 
has participated.
In any other cases, we manually resolve the inconsistency. 
If we cannot confidently assign a tag, we remove the address from the database.
When applying this procedure to the \ntags addresses, only 66 (0.03\%)
required manual inspection and we could disambiguate those, 
showing that this strategy is effective. 

Next, we obtain the MI clusters for all the tagged addresses. 
If a single tagged address appears in the cluster, 
we assign that tag to all other addresses in the cluster.
If multiple tagged addressed appear in the same cluster, 
we apply the following disambiguation process.
If a cluster has only one service tag, 
we assume this is a case of double ownership where the owner of the address 
is the service, and the beneficiary is the owner of the other tags.
If multiple service tags are present, we try to manually disambiguate them, 
e.g., checking if the two services are related to each other. 
For cases where no service tags are present, 
we compute the edit distance between their labels. 
If they are close, e.g., user:Metabank.ru and user:Metabank, 
we consider them aliases and merge them. 
Otherwise, we manually try to disambiguate them.
If manual disambiguation fails, we do not expand the tags to the cluster.

Applying MI clustering to the \ntags tagged addresses
produces \nclusters clusters comprising of \nctags addresses.
Of the \nclusters clusters, 817 (1.15\%) have more than one tag,
from which 368 have only miscabuse tags from Bitcoin Abuse
and 286 can be automatically solved
since they contain multiple tagged addresses of the same owner.
Of the remaining 163 clusters, we could manually resolve conflicts in 91
corresponding to 54 services with double-ownership (e.g. exchanges)
and 37 equivalent tags (e.g., usernames with two aliases).
We could not resolve 72 clusters, for which we do not propagate tags.
Of those, 66 were clusters without service tags
(63 of them with multiple Bitcoin-OTC user tags) and
6 clusters with more than one service tag
(4 of them with multiple mining tags).
An interesting case is one of the clusters with multiple service tags.
It is due to the special address 3J98t1, 
which corresponds to the empty string private key. 
Surprisingly, that address occasionally receives deposits possibly because 
it appears in many Bitcoin tutorials. 
Any Bitcoin user can spend funds from this address 
(since the private key is not really secret), and many users actually do, 
creating a MI cluster with 13M addresses. 
In summary, only 0.27\% of the 61K clusters required manual inspection,
and we could expand tags for 99.3\% of all tagged clusters,
showing that our strategy is effective.

\section{Exchange Detector}
\label{sec:classifier}

We build a binary ML classifier 
to identify if an address belongs to an exchange 
in three steps. 
First, we build a dataset of 108K labeled addresses. Then, we propose 42 features 
and use them to train different models.
Finally, we evaluate the accuracy of the models.

\begin{table}[t]
\centering
\scriptsize
\begin{tabular}{|l l r r r r|}
\hline
	\textbf{Class} & \textbf{Source} & \textbf{StAddr.} & \textbf{Entities} & \textbf{MI Clust.} & \textbf{Addr.} \\
\hline
	positive	& WalletExplorer &   167 & 117 & 151 & 54,265 \\ 	negative	& Bitcoin-OTC	 &   923 & 923 & 788 & 38,702 \\ 	negative	& SatoshiDice	 &    40 &   1 &  13 &  1,154 \\ 	negative	& Ransomware	 & 7,223 &  65 &  78 & 14,409 \\ \ignore{
	\textbf{Class} & \textbf{Source} & \textbf{Addr.} & \textbf{Entities} & \textbf{MI Clust.} & \textbf{MI Addr.} \\
\hline
	positive	& WalletExplorer &  167   & 117	& 151 & 28,875,775 \\ 	negative	& Bitcoin-OTC	 &  923   & 923	&  788 &     39,233 \\ 	negative	& SatoshiDice	 &   40   & 1	& 13 &      1,154 \\ 	negative	& Ransomware	 & 7,223   & 65	&  78 &     10,444 \\ }
\hline
\end{tabular}
\caption{Dataset of 108,530 addresses used for training and testing exchange classifier.}
\label{table:dataset}
\vspace{-0.5cm}
\end{table}

\paragraph{Dataset.}
To train and test the classifier we build a dataset of
108,530 labeled addresses, summarized in Table~\ref{table:dataset}. 
Half of them (positive class) are addresses
tagged as exchanges by WalletExplorer~\cite{walletexplorer}.
The other half (negative class) are non-exchange addresses
belonging to individuals, ransomware, and gambling sites.
Individuals correspond to Bitcoin-OTC~\cite{bitcoin-otc} users
with a registered Bitcoin address and GPG key, and 
having received ratings. 
To avoid abusive users we require that they have  
received less than 100 negative ratings and have an overall positive rating.
Ransomware addresses come from~\cite{ransomwareClouston} and
gambling addresses belong to SatoshiDice~\cite{satoshiDice}.
We obtain the MI clusters of those starting addresses and 
select additional addresses from the clusters until the dataset is balanced.
We use 80\% of the dataset for training and hold the remaining 20\% for testing.

\begin{table}[t]
\centering
\resizebox{\columnwidth}{!}{
\begin{tabular}{|l c l c l|}
\hline
	\textbf{Feature} & \textbf{New} & \textbf{Type} & \textbf{Class} & \textbf{Description} \\
\hline
			type		& \Y & Cat. & AD & Address type \\
	equiv\_addrs	& \Y & Int. & AD & Number of equivalent addresses \\
		lifetime	& \N & Int. & TS  & Total seconds from first TX to last \\
	timespan\_d	& \Y & Int. & TS  & Seconds from first deposit TX to last \\
	timespan\_w	& \Y & Int. & TS  & Seconds from first withdrawal TX to last \\
	activity	& \N & Int. & TS  & Number of days with at least one TX \\
	activity\_d	& \Y & Int. & TS  & Number of days with at least one deposit TX \\
	activity\_w	& \Y & Int. & TS  & Number of days with at least one withdrawal TX \\
	idle\_time	& \Y & Int. & TS  & Number of days without TXES \\
		daily\_d\_rate	& \Y & Float & TS  & Rate of deposits per day in its lifetime (tx/day) \\
	daily\_w\_rate	& \Y & Float & TS  & Rate of withdrawals per day in its lifetime (tx/day)\\
	yearly\_d\_txes	& \Y & Float & TS  & Average number of deposit TXES per year \\
	yearly\_w\_txes	& \Y & Float & TS  & Average number of withdrawal TXES per year \\
		balance		& \Y & Int. & TX & Balance to date, in satoshis \\
	deposited	& \N & Int. & TX & Total satoshis deposited/received into address \\
	withdrawn	& \N & Int. & TX & Total satoshis withdrawn/sent from address \\
	txes		& \N & Int. & TX & Total number of transactions \\
	txes\_out	& \N & Int. & TX & Number of TXES with this address as output \\
	txes\_in	& \N & Int. & TX & Number of TXES with this address as input \\
	addr\_as\_change& \Y & Float & TX & Ratio of TXES address used as input and output \\
	outputs		& \Y & Int. & TX & Total output slots in which the address appear \\
	inputs		& \Y & Int. & TX & Total input slots in which the address appear \\
	utxos		& \Y & Int. & TX & Number of unspent transaction outputs to date \\
	tx\_size\_mean	& \Y & Float & TX & Average total size (in bytes) of all its TXES \\
	tx\_weight\_mean& \Y & Float & TX & Average weight of TXES \\ 	tx\_fee\_mean	& \Y & Float & TX & Average fee paid in TXES \\
	ins\_age\_mean	& \Y & Float & TX & Average blocks between input and spent output \\
	coinbase	& \N & Int. & TX & Number of TXES that are coinbase \\
	coinjoin	& \Y & Int. & TX & Number of TXES that are coinjoin \\
	coinjoin\_out	& \Y & Int. & TX & Coinjoin TXES with this address as output \\
	coinjoin\_in	& \Y & Int. & TX & Coinjoin TXES with this address as input \\
		tx\_ratio	& \N & Float & TX & Number of input TXES divided by output TXES \\
	outs\_per\_tx	& \Y & Float & TX & Average number of output slots per TX \\
	ins\_per\_tx	& \Y & Float & TX & Average number of input slots per TX \\
	outs\_per\_out	& \Y & Float & TX & Average number of output slots per output TX \\
	ins\_per\_out	& \Y & Float & TX & Average number of input slots per output TX \\
	outs\_per\_in	& \N & Float & TX & Average number of output slots per input TX \\
	ins\_per\_in	& \N & Float & TX & Average number of input slots per input TX \\
		profit\_rate	& \Y & Float & TX & Total satoshis deposited over lifetime \\
	expense\_rate	& \Y & Float & TX & Total satoshis withdrawn over lifetime \\
	d\_per\_tx	& \Y & Float & TX & Total deposited amount per TX (satoshis/tx) \\
	w\_per\_tx	& \Y & Float & TX & Total withdrawn amount per TX (satoshis/tx) \\
					\hline
\end{tabular}
}
\caption{\small{Classifier features extracted per address.}}
\label{table:features}
\vspace{-0.4cm}
\end{table}

\paragraph{Model training.}
We extract 42 features for an address,
summarized in Table~\ref{table:features}.
Features can be grouped into three classes.
The 29 transaction (TX) features 
capture properties of the transactions where the address appears such as 
number of (all/deposit/withdrawal/CoinJoin) transactions, 
number of transactions where the address appears as input or output, and
statistics on the transferred amounts and paid fees. 
The 11 block timestamp (TS) features capture time properties 
such as address lifetime, time differences between noteworthy events 
(e.g., first and last deposits), and 
statistics on the daily rate of transactions.
Finally, two address (AD) features 
capture the address type (e.g., pubkeyhash, multisig) and 
the number of other addresses (e.g., multisig) built from the address.
Of the 42 features, 74\% (31) are novel and the rest have been used 
in previous works~\cite{ponziBartoletti,firstSun,DBLP:conf/hicss/HarlevYLMV18, evaluationLin, Jourdan2018CharacterizingEI}.
Features are normalized using z-score to have
zero mean and unit standard deviation.
To understand the importance of the different features, 
we perform a feature analysis using mutual information (MI) gain. 
The feature analysis shows that the most important features are those that 
capture the high volume of transactions over time by exchanges
The top 5 features and their MI gain are:
yearly\_d\_txes (0.1626), 
yearly\_w\_txes (0.1601),
activity\_w (0.1475)
txes\_in (0.1463)
and
daily\_w\_rate (0.1437).

\begin{table}[t]
\centering
\scriptsize
\begin{tabular}{|l | r r r | r r r|}
\cline{2-7}
	\multicolumn{1}{c|}{} & \multicolumn{3}{c|}{\textbf{Training}} & \multicolumn{3}{c|}{\textbf{Testing}} \\
\hline
	\textbf{Classifier} & \textbf{Prec.} & \textbf{Rec.} & \textbf{F1} & \textbf{Prec.} & \textbf{Rec.} & \textbf{F1} \\
\hline
Random Forest & 0.956 & 0.956 & 0.956 & 0.971 & 0.946 & 0.958 \\
Gradient Boosting & 0.949 & 0.948 & 0.948 & 0.965 & 0.935 & 0.950 \\
Decision Tree & 0.934 & 0.934 & 0.934 & 0.938 & 0.936 & 0.937 \\

\hline
\end{tabular}
\caption{\small{Classifier results using 
Stratified 5-fold Cross Validation on 80\% training data, and on 
20\% of unseen testing data.}}
\label{table:classification}
\vspace{-0.7cm}
\end{table}

We train three models using the 42 features and 80\% of the labeled dataset:
Decision Tree, Random Forest, and Gradient Boosting. 
We focus on tree-based models because they are easy to interpret.
The left side of Table~\ref{table:classification} 
shows the precision, recall, and F1 score 
for each model, using a stratified 5-fold cross validation training.
The results show that Random Forest achieves the best F1 score of 0.956.

\paragraph{Testing.}
We evaluate the models to make predictions on unseen data
by using the 20\% of the dataset reserved for testing.
The results, on the right side of Table \ref{table:classification},
show that the best model is again Random Forest with a F1 of 0.958, 
so we use that model throughout the explorations 
with the following hyper-parameters:
600 trees, maximum depth of 40, and prediction threshold of 0.5.
Figure~\ref{fig:roc} in the Appendix shows the classifier's ROC curve.

\begin{table*}
\centering
\scriptsize
\begin{tabular}{| c| l | c | r r r | r r r r | r r | r | r|}
\cline{4-12}
	\multicolumn{2}{c}{} & \multicolumn{1}{c|}{\textbf{}} & \multicolumn{3}{c|}{\textbf{Seeds}} & \multicolumn{4}{c|}{\textbf{Graph}} & \multicolumn{2}{c|}{\textbf{Classification}} & \multicolumn{1}{c}{} \\
\hline
	 \textbf{Class} & \textbf{Operation} 
  & \textbf{SDD}
	& \textbf{All} & \textbf{OW} & \textbf{Exp.}
	& \textbf{Comp.} & \textbf{Addr} & \textbf{Txes} & \textbf{Unexp.}
	& \textbf{Tagged}
	& \textbf{Exch.} 
  & \textbf{Relat.}
  & \textbf{Time}\\
\hline
	\multirow{4}{*}{\rotatebox{90}{C\&C}}  
	&cerber		 & \Y &     4 &  0 &     4 &  1 &   608 (73.7\%) &   1,050 (86.0\%) &     1 &    12 &   394 &   1 (0) &   1'26" \\
	&glupteba	 & \Y &     3 &  0 &     3 &  3 &    195 (17.4\%) &      167 (24.6\%) &     5 &    14 &    83 &   0 (0) &   1'02" \\
	&pony		 & \Y &     4 &  0 &     4 &  2 & 1,154 (2.3\%) &   4,252 (36.4\%) &    72 &   379 &   188 &   1 (0) &  5'57" \\
	&skidmap	 & \Y &     1 &  1 &     1 &  1 &   132 (0.8\%) &     107 (11.2\%) &     5 &    37 &    7 &   5 (1) &   4'04" \\ \hline
	\multirow{4}{*}[-1.5cm]{\rotatebox{90}{Clipper}}
	&aggah		 & \N &     1 &  0 &     1 &  1 &   54 (13.0\%)  &      44 (18.2\%) &     6 &     7 &    19 &   1 (1) &   1'04" \\
	&androidclipper	 & \N &     1 &  1 &     0 &  1 &       1 (-) &         0 (-) &     0 &     1 &     0 &   1 (1) &       - \\ 	&azorult	 & \N &     1 &  1 &     0 &  1 &       1 (-) &         0 (-) &     0 &     1 &     0 &   1 (1) &       - \\ 	&bitcoingrabber	 & \N &     1 &  0 &     1 &  1 &   504 (18.7\%) &     650 (46.9\%) &    32 &   203 &   116 &   0 (5) &   3'19" \\
	&casbaneiro	 & \N &     1 &  0 &     1 &  1 &   125 (47.2\%) &     162 (59.3\%) &     5 &    48 &    55 &   4 (4) &   0'51" \\
	&clipbanker	 & \N &     1 &  1 &     0 &  1 &       1 (-) &         0 (-) &     0 &     1 &     0 &   1 (1) &       - \\ 	&clipsa  	 & \N & 9,412 &  0 &   375 & 25 & 4,655 (14.6\%) &   5,756 (14.6\%) &   120 & 1,339 & 2,471 &   6 (6) &  25'18" \\
	&cliptomaner	 & \N &     1 &  1 &     0 &  1 &       1 (-) &         0 (-) &     0 &     1 &     0 &   1 (1) &       - \\ 	&cryptoshuffler  & \N &     1 &  0 &     1 &  1 & 3,253 (15.6\%) &   7,744 (30.1\%) &   220 &   966 & 1,279 &  11 (12) &  42'00" \\
	&masad		 & \N &     2 &  0 &     2 &  1 & 3,882 (12.1\%) &   7,389 (20.0\%) &   182 & 1,366 & 1,404 &  11 (11) &  28'40" \\
	&mekotio	 & \N &     3 &  0 &     2 &  1 & 2,180 (6.7\%) &   4,031 (9.6\%) &   71 &   600 & 1,125 &   6 (17) &  21'08" \\
	&mekotion40	 & \N &     3 &  0 &     3 &  3 & 1,153 (12.6\%) &   1,152 (27.0\%) &    30 &   507 &   338 &   6 (8) &  24'25" \\
	&mispadu	 & \N &     1 &  1 &     0 &  1 &       1 (-) &         0 (-) &     0 &     1 &     0 &   1 (1) &       - \\ 	&mrpr0gr4mmer	 & \N &     2 &  0 &     2 &  1 &   837 (68.9\%) &   1,704 (92.1\%) &    19 &   321 &   350 &   2 (2) &   2'18" \\
	&n40		 & \N &     1 &  1 &     0 &  1 &       1 (-) &         0 (-) &     0 &     1 &     0 &   1 (1) &       - \\ 	&phorpiex-tldr	 & \N &    15 & 14 &     1 &  1 &   203 (5.4\%) &     107 (51.4\%) &    14 &   109 &    62 &   3 (3) &   2'16" \\ 	&phorpiex-trik	 & \N &     8 &  1 &     6 &  4 &   282 (36.5\%) &     539 (35.0\%) &     2 &    19 &   116 &   5 (8) &   1'14" \\ 	&predatorthethief& \N &     1 &  0 &     1 &  1 & 2,702 (15.2\%) &   5,039 (32.3\%) &   157 & 1,100 & 947 &   8 (9) &  38'22" \\
	&protonbot	 & \N &     2 &  2 &     0 &  1 &       2 (-) &         0 (-) &     0 &     2 &     0 &   2 (2) &       - \\ \hline
	\multirow{4}{*}[-0.25cm]{\rotatebox{90}{Ransomware}}
    &cryptotorlocker2015 & \N &     1 &  0 &     1 &  1 &    336 (43.8\%) &  3,695 (57.8\%) &    31 &    103 &   125 &  11 (8) & 122'02" \\
	&cryptxxx	 & \N &     1 &  0 &     1 &  1 &  4,015 (15.5\%) & 10,087 (18.2\%) &    43 & 1,607 & 984 &  13 (11) &  31'29" \\
	&dmalocker	 & \N &    11 &  0 &    11 &  1 &  1,631 (29.6\%) &  4,855 (28.3\%) &    25 &   525 &   531 &   8 (6) &   3'44" \\
	&globeimposter	 & \N &     3 &  0 &     1 &  1 & 13,821 (4.3\%) & 62,570 (8.4\%) & 2,871 & 3,614 & 4,737 &   9 (7) & 112'42" \\
	&locky		 & \N & 7,076 &  0 & 7,074 &  1 &  9,546 (83.7\%) & 22,688 (51.5\%) &    89 & 7,303 &   715 &  15 (15) &  22'59" \\
	&samsam		 & \N &    45 &  0 &    20 & 10 &  1,020 (29.7\%) &  1,108 (48.1\%) &    15 &   184 &   300 &   6 (6) &   2'52" \\
	&wannacry	 & \N &     6 &  0 &     5 &  2 &     38 (78.9\%) &    531 (97.9\%) &     0 &     18 &    10 &   2 (2) &   0'25" \\
\hline
\end{tabular}

\ignore{
\begin{table*}
\centering
\scriptsize
\begin{tabular}{| c| l | c | r r r | r r r r | r r | r | r|}
\cline{4-12}
	\multicolumn{2}{c}{} & \multicolumn{1}{c|}{\textbf{}} & \multicolumn{3}{c|}{\textbf{Seeds}} & \multicolumn{4}{c|}{\textbf{Graph}} & \multicolumn{2}{c|}{\textbf{Classification}} & \multicolumn{1}{c}{} \\
\hline
	 \textbf{Class} & \textbf{Operation} 
  & \textbf{SDD}
	& \textbf{All} & \textbf{OW} & \textbf{Exp.}
	& \textbf{Comp.} & \textbf{Addr} & \textbf{Txes} & \textbf{Unexp.}
	& \textbf{Tagged}
	& \textbf{Exch.} 
  & \textbf{Relat.}
  & \textbf{Time}\\
\hline
	\multirow{4}{*}{\rotatebox{90}{C\&C}}  
	&cerber		 & \Y &    4 &  0 &    4 &  1 &  767 (59.1\%) & 1,204 (75.4\%) &   5 &   23 &  516 & 1 (0) & 4.2m \\
	&pony		 & \Y &    4 &  0 &    4 &  2 & 1,159 (2.3\%) & 4,271 (36.2\%) &  87 &  363 &  492 & 0 (0) & 19.1m \\
	&skidmap	 & \Y &    1 &  1 &    1 &  1 &  132 (0.8\%) &  107 (11.2\%) &   7 &   33 &   16 & 3 (0) & 5.6m \\ \hline
	\multirow{4}{*}[-1.5cm]{\rotatebox{90}{Clipper}}
	&aggah		 & \N &    1 &  0 &    1 &  1 &   96 (7.3\%) &   63 (12.7\%) &  15 &   23 &   43 & 1 (1) & 2.8m \\
	&androidclipper	 & \N &    1 &  1 &    0 &  1 &    1 (-) &    0 (-) &   0 &    1 &    0 & 1 (1) & - \\ 	&azorult	 & \N &    1 &  1 &    0 &  1 &    1 (-) &    0 (-) &   0 &    0 &    1 & 0 (0) & - \\ 	&bitcoingrabber	 & \N &    1 &  0 &    1 &  1 &  504 (18.7\%) &  650 (46.9\%) &  48 &  170 &  222 & 0 (2) & 9.9m \\
	&casbaneiro	 & \N &    1 &  0 &    1 &  1 &  131 (45.0\%) &  182 (52.7\%) &   9 &   50 &   71 & 1 (1) & 12.0m \\
	&clipbanker	 & \N &    1 &  1 &    0 &  1 &    1 (-) &    0 (-) &   0 &    1 &    0 & 1 (1) & - \\ 	&clipsa  	 & \N & 9,412 &  0 &  375 & 25 & 7,849 (8.7\%) & 9,659 (8.7\%) & 238 & 2,262 & 5,558 & 4 (5) & 138.1m \\
	&cliptomaner	 & \N &    1 &  1 &    0 &  1 &    1 (-) &    0 (-) &   0 &    0 &    1 & 0 (0) & - \\ 	&cryptoshuffler  & \N &    1 &  0 &    1 &  1 & 3,333 (15.4\%) & 7,749 (30.1\%) & 230 &  889 & 1,680 & 9 (9) & 60.4m \\
	&masad		 & \N &    2 &  0 &    2 &  1 & 5,931 (10.7\%) & 9,942 (21.5\%) & 410 & 2,250 & 3,667 & 3 (3) & 119.9m \\
	&mekotio	 & \N &    3 &  0 &    2 &  1 & 2,180 (6.7\%) & 4,149 (9.3\%) & 116 &  589 & 1,446 & 5 (8) & 47.1m \\
	&mekotion40	 & \N &    3 &  0 &    3 &  3 & 1,204 (12.0\%) & 1,174 (26.5\%) &  45 &  494 &  423 & 4 (5) & 39.3m \\
	&mispadu	 & \N &    1 &  1 &    0 &  1 &    1 (-) &    0 (-) &   0 &    1 &    0 & 1 (1) & - \\ 	&mrpr0gr4mmer	 & \N &    2 &  0 &    2 &  1 & 1,052 (56.9\%) & 1,806 (88.7\%) &  46 &  378 &  513 & 1 (1) & 10.7m \\
	&n40		 & \N &    1 &  1 &    0 &  1 &    1 (-) &    0 (-) &   0 &    1 &    0 & 1 (1) & - \\ 	&phorpiex-tldr	 & \N &   15 & 14 &    1 &  1 &  203 (5.4\%) &  107 (51.4\%) &  28 &  108 &   90 & 2 (2) & 7.4m \\ 	&phorpiex-trik	 & \N &    8 &  1 &    6 &  4 &  406 (25.4\%) &  878 (21.5\%) &   6 &   22 &  178 & 3 (6) & 2.2m \\ 	&predatorthethief& \N &    1 &  0 &    1 &  1 & 2,702 (15.2\%) & 5,065 (32.1\%) & 204 & 1,074 & 1,403 & 2 (2) & 65.0m \\
	&protonbot	 & \N &    2 &  2 &    0 &  1 &    2 (-) &    0 (-) &   0 &    2 &    0 & 2 (2) & - \\ \hline
	\multirow{4}{*}[-0.25cm]{\rotatebox{90}{Ransomware}}
    &cryptotorlocker2015 & \N &    1 &  0 &    1 &  1 &  336 (43.8\%) & 4,212 (61.2\%) &  36 &   94 &  125 & 11 (8) & 139.1m \\
	&cryptxxx	 & \N &    1 &  0 &    1 &  1 & 4,019 (15.5\%) & 10,157 (18.1\%) &  48 & 1,590 & 1,009 & 10 (7) & 39.7m \\
	&dmalocker	 & \N &   11 &  0 &   11 &  1 & 2,206 (26.8\%) & 8,773 (39.8\%) &  53 &  275 &  762 & 7 (4) & 10.4m \\
	&globeimposter	 & \N &    3 &  0 &    1 &  1 & 31,152 (2.8\%) & 123,124 (4.8\%) & 4,079 & 8,762 & 14,494 & 8 (4) & 26.2h \\
	&locky		 & \N & 7,076 &  0 & 7,074 &  1 & 9,579 (83.5\%) & 22,973 (51.4\%) &  92 & 7,297 &  767 & 11 (12) & 37.3m \\
	&samsam		 & \N &   45 &  0 &   20 & 10 & 1,023 (29.6\%) & 1,112 (47.9\%) &  15 &  182 &  310 & 5 (5) & 5.4m \\
	&wannacry	 & \N &    6 &  0 &    5 &  2 &  308 (11.0\%) &  532 (98.5\%) &   1 &    8 &  290 & 0 (0) & 2.7m \\
\hline
\end{tabular}
}

\ignore{
\begin{table*}
\centering
\scriptsize
\begin{tabular}{| c| l | c | r r r | r r r r | r r | r|}
\cline{4-12}
	\multicolumn{2}{c}{} & \multicolumn{1}{c|}{\textbf{}} & \multicolumn{3}{c|}{\textbf{Seeds}} & \multicolumn{4}{c|}{\textbf{Graph}} & \multicolumn{2}{c|}{\textbf{Classification}} & \multicolumn{1}{c}{} \\
\hline
	 \textbf{Class} & \textbf{Operation} 
  & \textbf{SDD}
	& \textbf{All} & \textbf{OW} & \textbf{Exp.}
	& \textbf{Comp.} & \textbf{Addr} & \textbf{Txes} & \textbf{Unexp.}
	& \textbf{Tagged}
	& \textbf{Exchanges} & \textbf{Runtime}\\
\hline
	\multirow{4}{*}{\rotatebox{90}{C\&C}}  
	&cerber		 & \Y &    4 &  0 &    4 &  1 &  767 & 1,204 &   5 &   23 &  516 & 4m11,662s \\
	&pony		 & \Y &    4 &  0 &    4 &  2 & 1,159 & 4,271 &  87 &  363 &  492 & 19m4,359s \\
	&skidmap	 & \Y &    1 &  1 &    1 &  1 &  132 &  107 &   7 &   33 &   16 & 5m37,554s \\ \hline
	\multirow{4}{*}[-1.5cm]{\rotatebox{90}{Clipper}}
	&aggah		 & \N &    1 &  0 &    1 &  1 &   96 &   63 &  15 &   23 &   43 & 2m47,830s \\
	&androidclipper	 & \N &    1 &  1 &    0 &  1 &    1 &    0 &   0 &    1 &    0 & - \\ 	&azorult	 & \N &    1 &  1 &    0 &  1 &    1 &    0 &   0 &    0 &    1 & - \\ 	&bitcoingrabber	 & \N &    1 &  0 &    1 &  1 &  504 &  650 &  48 &  170 &  222 & 9m57,435s \\
	&casbaneiro	 & \N &    1 &  0 &    1 &  1 &  131 &  182 &   9 &   50 &   71 & 12m0,266s \\
	&clipbanker	 & \N &    1 &  1 &    0 &  1 &    1 &    0 &   0 &    1 &    0 & - \\ 	&clipsa  	 & \N & 9,412 &  0 &  375 & 25 & 7,849 & 9,659 & 238 & 2,262 & 5,558 & 138m6,446s \\
	&cliptomaner	 & \N &    1 &  1 &    0 &  1 &    1 &    0 &   0 &    0 &    1 & - \\ 	&cryptoshuffler  & \N &    1 &  0 &    1 &  1 & 3,333 & 7,749 & 230 &  889 & 1,680 & 60m22,250s \\
	&masad		 & \N &    2 &  0 &    2 &  1 & 5,931 & 9,942 & 410 & 2,250 & 3,667 & 119m57,446s \\
	&mekotio	 & \N &    3 &  0 &    2 &  1 & 2,180 & 4,149 & 116 &  589 & 1,446 & 47m6,552s \\
	&mekotion40	 & \N &    3 &  0 &    3 &  3 & 1,204 & 1,174 &  45 &  494 &  423 & 39m16,242s \\
	&mispadu	 & \N &    1 &  1 &    0 &  1 &    1 &    0 &   0 &    1 &    0 & - \\ 	&mrpr0gr4mmer	 & \N &    2 &  0 &    2 &  1 & 1,052 & 1,806 &  46 &  378 &  513 & 10m45,326s \\
	&n40		 & \N &    1 &  1 &    0 &  1 &    1 &    0 &   0 &    1 &    0 & - \\ 	&phorpiex-tldr	 & \N &   15 & 14 &    1 &  1 &  203 &  107 &  28 &  108 &   90 & 7m24,693s \\ 	&phorpiex-trik	 & \N &    8 &  1 &    6 &  4 &  406 &  878 &   6 &   22 &  178 & 2m14,104s \\ 	&predatorthethief& \N &    1 &  0 &    1 &  1 & 2,702 & 5,065 & 204 & 1,074 & 1,403 & 65m3,391s \\
	&protonbot	 & \N &    2 &  2 &    0 &  1 &    2 &    0 &   0 &    2 &    0 & - \\ \hline
	\multirow{4}{*}[-0.25cm]{\rotatebox{90}{Ransomware}}
    &cryptotorlocker2015 & \N &    1 &  0 &    1 &  1 &  336 & 4,212 &  36 &   94 &  125 & 139m8,751s \\
	&cryptxxx	 & \N &    1 &  0 &    1 &  1 & 4,019 & 10,157 &  48 & 1,590 & 1,009 & 39m40,304s \\
	&dmalocker	 & \N &   11 &  0 &   11 &  1 & 2,206 & 8,773 &  53 &  275 &  762 & 10m22,057s \\
	&globeimposter	 & \N &    3 &  0 &    1 &  1 & 31,152 & 123,124 & 4,079 & 8,762 & 14,494 & 1572m56,678s \\
	&locky		 & \N & 7,076 &  0 & 7,074 &  1 & 9,579 & 22,973 &  92 & 7,297 &  767 & 37m19,895s \\
	&samsam		 & \N &   45 &  0 &   20 & 10 & 1,023 & 1,112 &  15 &  182 &  310 & 5m23,499s \\
	&wannacry	 & \N &    6 &  0 &    5 &  2 &  308 &  532 &   1 &    8 &  290 & 2m45,626s \\
\hline
\end{tabular}
}

\ignore{
\begin{tabular}{|l r c r r r r r r r r r r r r r|}
\hline
	& \multicolumn{5}{c}{\textbf{Config}} & \multicolumn{2}{c}{\textbf{Total}} & \multicolumn{2}{c}{\textbf{Op.Addrs.}} & \multicolumn{3}{c}{\textbf{Tags}} & \multicolumn{3}{c|}{\textbf{Classif.}} \\
\hline
	\textbf{Operation}
	& \textbf{Steps} & \textbf{SDD} & \textbf{predTH} & \textbf{maxN} & \textbf{maxT}
	& \textbf{Addrs} & \textbf{Txes}
	& \textbf{Seeds} & \textbf{OP}
	& \textbf{Exchanges} & \textbf{Serv.} & \textbf{Other}
	& \textbf{Exchanges} & \textbf{NE} & \textbf{Unexp.} \\
\hline
	cerber		 & 3 & \N & 0.5 &  500 &   0 &  736 &  995 & 1 & 134 &  2 & 0 &   9 & 572 & 164 &  0 \\
	pony		 & 4 & \N & 0.5 & 1250 &   0 &  380 & 2622 & 1 & 245 &  1 & 0 &  22 & 107 & 271 &  2 \\
	pony		 & 4 & \N & 0.7 & 1250 &   0 &  475 & 3471 & 1 & 315 &  2 & 0 &  25 &  56 & 416 &  3 \\
	pony new	 & 4 & \N & 0.5 & 1250 &   0 &   25 &   76 & 1 &  15 &  0 & 0 &   1 &   9 &  16 &  0 \\
	skidmap		 & 4 & \N & 0.5 &  200 &  50 &   99 & 3929 & 1 &   0 &  7 & 1 &   8 &  23 &  74 &  2 \\ 	aggah		 & 6 & \Y & 0.5 &  500 &   0 &   47 &   94 & 1 &   0 &  1 & 0 &   5 &  22 &  24 &  1 \\
	bitcoingrabber	 & 2 & \Y & 0.5 &  200 &  50 &  314 &  914 & 1 &   0 & 18 & 0 &  49 & 125 & 181 &  8 \\
	phorpiex-tldr	 & 3 & \Y & 0.5 &  200 &   0 &  203 &  107 & 1 &   0 &  9 & 0 & 100 & 114 &  41 & 48 \\
	phorpiex-trik	 & 3 & \Y & 0.5 &  500 & 200 &  284 &  541 & 6 &   3 &  5 & 1 &  13 & 119 & 161 &  4 \\
	mekotion40	 & 3 & \Y & 0.5 &  500 & 200 & 1017 & 2393 & 3 &   4 &206 &25 &  92 & 431 & 568 & 18 \\
	mekotio		 & 3 & \Y & 0.5 &  500 & 200 & 2181 & 8207 & 2 &   6 &160 & 2 & 433 &1509 & 642 & 30 \\
	masad		 & 2 & \Y & 0.5 & 2000 &   - & 2170 & 4140 & 2 &   0 &118 & 5 & 890 &1268 & 807 & 95 \\
	predatorthethief & 2 & \Y & 0.5 & 2000 &   - & 2276 & 3751 & 1 &  38 &281 & 0 & 396 &1448 & 705 &123 \\
	clipsa  	 & 3 & \Y & 0.5 &  500 & 200 &10211 &20154 &375& 203 &179 &19 &2056 &7803 &2165 &243 \\
	dmalocker	 & 3 & \Y & 0.5 & 2000 &   - & 1629 & 4874 &10 & 155 &318 &16 & 187 & 536 &1068 & 25 \\
	wannacry	 & 3 & \Y & 0.5 & 2000 &   - &   55 &  559 & 6 &   0 &  1 & 0 &   7 &  27 &  27 &  1 \\
	locky		 & 3 & \Y & 0.5 & 2000 &   - & 9579 &22973 &7074&7074&142 &44 &7137 & 741 &8746 & 92 \\
	samsam		 & 3 & \Y & 0.5 & 2000 &   - & 1023 & 1112 &20 &  68 &102 & 3 &  77 & 310 & 697 & 16 \\
	cryptxxx	 & 3 & \Y & 0.5 & 2000 &   - & 2896 & 5699 & 1 &1464 & 25 & 0 &1469 & 491 &2394 & 11 \\
	globeimposter	 & 3 & \Y & 0.5 & 2000 &   - & 2448 &19415 & 1 &   0 &401 &40 &  39 & 905 &1394 &149 \\
\hline
\end{tabular}
}
\caption{\Explorations for the \numoperations malware families. In parenthesis, forward-only exploration results for comparison: fraction of addresses and transactions and total number of relations present in the forward-only graph.
}
\label{tab:graphs}
\end{table*}

\section{\EXploration Evaluation}
\label{sec:evaluation}

We evaluate \tool on \numoperations malware families: 
\numoperationsmalware families that use Bitcoin for C\&C, \numoperationsclipper clippers that replace addresses in the OS clipboard 
of the infected host with an address controlled by the malware operators, and 
\numoperationsransomware ransomware.
We compare \exploration with a baseline forward-only exploration
where we turn off the backwards exploration step in \tool,
but allow forward-only exploration to benefit from our 
exchange classifier and tag database.
We use this baseline because prior works implement different 
versions of forward-only exploration that vary, among others, 
in the number of steps or addresses explored,
whether MI clustering is used, and
what tags or other heuristics are used to filter unrelated addresses. 
For each family, we run two explorations: back-and-forth and forward-only.
Table~\ref{tab:graphs} summarizes the \explorations and the delta with 
the forward-only explorations (in parenthesis).
Full forward-only results are in Table~\ref{tab:graphs-skipback} 
in the Appendix.

For clipper and ransomware families \exploration is configured to avoid 
exploring backwards on the seeds since addresses depositing into the seeds 
belong to victims (SDD column). 
Note however, that \exploration will still explore backwards after the first 
forward step from the seeds.
The seed information is broken into three columns. 
\emph{All} is the total addresses in our tag database for the family. 
\emph{OW} captures how many of those seeds are identified 
as belonging to online wallets.
Those seeds are not explored forward as that would explore 
the transactions of the service holding the online wallet, 
but they can still be explored backwards if they do not belong to a clipper or 
ransomware (i.e., skidmap). 
\emph{Exp.} captures the number of addresses effectively explored, 
after removing addresses in online wallets and addresses without transactions.
Some families like \tag{clipsa} and \tag{locky} have thousands of seeds.
In the case of \tag{clipsa}, when replacing the address in the clipboard, 
this clipper selects a visually-similar address 
from an address database,
so that the replacement has a higher chance of not
being noticed by the victim.
Since the address database ships with the malware, researchers have published
its contents~\cite{clipsa}.
However, only 375 (4\%) of clipsa addresses have transactions and thus 
are effectively used in the exploration. 
Those 375 addresses collect nearly 18 BTC (\$277K).
Other families like n40 collect even larger 
BTC amounts (27.4 BTC, \$187K)
using a single replacement address.
Thus, despite their simplicity,
clippers can produce significant gains for their operators.
For the interested reader, Table~\ref{tab:clippers} in the Appendix 
shows the total amounts received by seed addresses.

An important observation is that 24 seeds are online wallets 
whose owner is an exchange and whose beneficiary is the malware operator 
(or a money mule).
Nearly all of those (23/24) are replacement addresses from 9 clipper families, 
the other belonging to the \tag{skidmap} crypto miner.
Thus, replacement addresses hardcoded in clipper samples are often hosted 
in exchanges. 
It is surprising that the abusive nature of those addresses 
goes unnoticed for the exchanges, 
as they are often mentioned in public abuse reports 
by the victims, as well as in analyses by security vendors. 
Note that this does not happen for addresses in ransom notes, 
possibly because of ransomware's higher visibility.
Based on this observation, we discuss in Section~\ref{sec:discussion} 
a defense that would examine public reports for malicious addresses, 
identify online wallets, and (publicly) report them to exchanges,  
so that it cannot be claimed that their abusive nature was not known.

The middle section of the table details
the output graphs. 
For 7 clipper families all seeds are in online wallets. 
Thus, no exploration is performed for them and 
the graph contains only the seeds.
For the remaining 23 families, 69\% contain a single component.
Each component contains at least one seed.
Thus, multiple components indicate that some seeds were not connected to 
the rest of the seeds. 
The graphs for those 23 families contain an average of 2,274 addresses,
with a maximum of 13.8K for \tag{globeimposter}. 
Unexplored addresses are those not processed before the 
exploration limit was reached.
The numbers in parenthesis show the fraction of addresses and 
transactions in the forward-only graph compared to the back-and-forth graph.
On average, the back-and-forth graph is three times larger 
(3.6x addresses, 2.6x transactions) offering a more 
complete view of an operation. 
In Sections~\ref{sec:discovery} and~\ref{sec:relations} 
we show that the back-and-forth graphs 
contain critical information not available in the forward-only graphs 
such as previously unknown C\&C signaling addresses for Cerber and Pony,
unreported relations between malware families (e.g. DMALocker and Slave),
and new attribution points (Pony, Skidmap, MrPr0gr4mmer, Phorpiextldr).

The right section of the table shows the number of addresses tagged 
and those without a tag identified as exchanges by the ML classifier.
It also shows the total number of relations found in the back-and-forth and 
forward-only (in parenthesis) explorations.
Surprisingly, forward-only exploration finds more relations (150) than 
back-and-forth exploration (142). 
The additional relations are FPs introduced by the exchange 
classifier failing to identify low-volume exchange addresses.
Figure~\ref{fig:singclassiffp} illustrates this.
Step 1 in the exploration finds address $A_c$, belonging to MI cluster $c$, 
which has no tag and is not identified as an exchange.
If \exploration is used, at step 2 $B_c$ and $F_m$ are introduced.
Address $F_m$ has an underground market tag, 
while address $B_c$ is classified as an exchange.
Now, our platform marks $A_c$ also 
as an exchange because it belongs to the same cluster $c$ as exchange 
address $B_c$, and removes paths starting from $A_c$ introduced in step 2.
Thus, the \exploration graph corresponds to the step 1 box and contains
a relation with exchange $A_c$.
However, if forward-only exploration is used, 
then step 2 only discovers the market address $F_m$. 
The exchange address $B_c$ is not discovered because it would require 
exploring backwards from $A_c$. 
Thus, the forward-only exploration includes the path from the seed to the 
market, which is a false relation introduced by the exchange address $A_c$.
While this issue can affect both explorations, 
it affects more often forward-only exploration as 
the probability of finding addresses in the same exchange cluster is smaller.

\begin{figure}[t]
  \centering
    \includegraphics{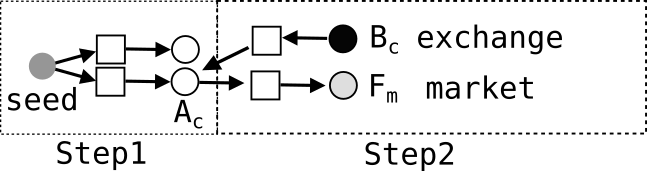}
  \vspace{-0.2cm}
  \caption{False positive example.}
  \label{fig:singclassiffp}
  \vspace{-0.45cm}
\end{figure}

The rightmost column captures the exploration runtime.
It ranges from a few seconds up to 2 hours,
with an average of 22 minutes per operation.
Forward-only exploration is faster with an average runtime of 7 minutes due
to the fewer transactions it analyzes.
However, Sections~\ref{sec:discovery} and~\ref{sec:relations} 
show forward-only exploration misses important relations that can only be 
identified by exploring backwards.

Next, we evaluate the accuracy of the produced graphs in 
Section~\ref{sec:discovery} and the impact of the exchange classifier in 
Section~\ref{sec:eval_classifier}.

\subsection{Evaluating Address Discovery}
\label{sec:discovery}

Evaluating the accuracy of the produced graphs is challenging as
we lack ground truth on the malicious operations.
However, for the four families that use the Bitcoin blockchain for C\&C
we are able to build operation oracles that can identify 
previously unknown C\&C signaling addresses in the output graph, 
demonstrating the benefits of \exploration.

The use of the Bitcoin blockchain for C\&C follows a similar pattern for 
the four families. 
Samples have a hardcoded signaling address. 
Upon infection, the sample queries a external Bitcoin monitoring service
(e.g.,~\cite{bitcoinExplorer})
to obtain the most recent transactions involving the signaling address.
Pony and Skidmap use the same signaling method that encodes 
C\&C IP addresses in the value of two consecutive deposits into the 
signaling address~\cite{pony}.
Cerber encodes C\&C domains in the value of a single transaction 
from the signaling address to a temporary address that then 
returns the funds~\cite{cerber}. 
And, Glupteba stores an encrypted C\&C domain 
in the blockchain using a transaction from the signaling address with 
an OP\_RETURN output~\cite{gluptebaTrendMicro}.
In all cases, the signaling transactions create a distinct pattern 
for which we build an operation oracle. 
The oracle determines if a given address is a signaling address 
for the family.
An exploration is performed from seeds obtained from previous reports 
and the operation oracle is applied to each address in the graph to 
identify previously unknown signaling addresses.

\begin{table}[t]
\centering
\scriptsize
\begin{tabular}{|l c r l r l|}
\hline
	\textbf{Address} & \textbf{In} & \textbf{Step} & \textbf{Found} & \textbf{FQDNs} & \textbf{Signaling Period}\\
\hline
17gd1m & \cite{cerber} & 0 & Seed & 254 & 2016-07-26 $\rightarrow$ 2017-06-13 \\ 1HTDy9 & \cite{cerber} & 1 & MI   & 128 & 2017-02-13 $\rightarrow$ 2017-07-23 \\ 1ML94w & \cite{cerber} & 1 & Forw &  34 & 2017-06-11 $\rightarrow$ 2017-09-16 \\ 1CpTCV & \cite{cerber} & 1 & MI   &   9 & 2017-08-31 $\rightarrow$ 2017-09-22 \\ 1GcnsL & \N & 1 & MI   &  17 & 2017-07-27 $\rightarrow$ 2017-09-22 \\ 14ru2h & \N & 2 & Back &   4 & 2017-01-21 $\rightarrow$ 2017-02-13 \\ 1DiRvu & \N & 2 & Back &   4 & 2017-03-12 $\rightarrow$ 2017-04-21 \\ \hline
\end{tabular}
\caption{Cerber signaling addresses. Exploring from 4 seeds, 3 previously unknown signaling addresses were discovered.}
\label{tab:cerber}
\vspace{-0.5cm}
\end{table}

\paragraph{Cerber.}
The Cerber ransomware was the first malware to use the Bitcoin blockchain 
for C\&C between July 2016 and late 2017 when the operators were arrested~\cite{cerberTakedown}.
The Cerber oracle captures that 
a signaling round creates cyclic transactions where a 
signaling address sends $x$ BTCs to a temporary address 
(only used for one signaling round)
that then, within minutes, returns 
$x-\textit{fee}$ to the signaling address, where \textit{fee} is the 
Bitcoin transaction fee (i.e., $0.001$ BTC). 
The first six characters of the temporary address correspond to a 
DNS domain on a specific TLD hardcoded in the sample.
The oracle is detailed in Algorithm~\ref{alg:cerber} in the Appendix.
Given an address, it iterates over its withdrawal transactions (line 2).
For each transaction, it performs five checks. 
First, it checks the transaction has a single input and a 
single output address (line 4).
Then it verifies that the recipient is a temporary address, 
i.e., has a single deposit transaction and a 
single withdrawal transaction (line 8).
Next, it validates that the recipient returns the received funds to the sender
by checking that the withdrawal transaction
has just one input and one output (line 12),
that its output is the input address (line 14), and
that it returns the value it received minus the transaction fee (line 16).

The Cerber exploration 
starts from four signaling addresses 
mentioned in~\cite{cerber} and identifies three previously unknown 
signaling addresses. 
All seven addresses are detailed in Table~\ref{tab:cerber}. 
For each signaling address, the table shows 
whether it was known,
the exploration step at which it was discovered,
the method used to discover it, 
the number of domains it signaled, and
the time period in which it signaled.
The three signaling addresses discovered are not mentioned in~\cite{cerber}, 
although they were active simultaneously with other seeds. 
One is identified by MI clustering,
but identifying the other two required exploring backwards.
This highlights how \exploration is instrumental to identify new addresses.

\begin{table}[t]
\centering
\scriptsize
\begin{tabular}{|l l r l r l|}
\hline
	\textbf{Address} & \textbf{In} & \textbf{Step} & \textbf{Found} & \textbf{IPs} & \textbf{Signaling Period} \\
\hline
	1BkeGq & \cite{pony}	& 0 & Seed & 237 & 2019-08-28 $\rightarrow$ 2021-01-13 \\ 	1CeLgF & \cite{pony}	& 0 & Seed &  99  & 2019-08-27 $\rightarrow$ 2019-12-11 \\ 	19hi8B & \cite{pony}	& 0 & Seed & 336 & 2019-12-10 $\rightarrow$ 2021-01-13 \\ 	1N9ALZ & \cite{ponyPeppermalware} & 0 & Seed & 19 & 2019-10-03 $\rightarrow$ 2020-08-06 \\ 	bc1qwl & \N		& 2 & Forw &   2   & 2019-07-24 $\rightarrow$ 2019-07-24 \\ 	1NL8QT & \N		& 2 & Forw &   3   & 2020-03-03 $\rightarrow$ 2020-07-22 \\ 	1GUghN & \N		& 2 & Back &   6   & 2020-03-03 $\rightarrow$ 2020-07-11 \\ 	bc1qh9 & \N		& 2 & Back &  24  & 2019-06-09 $\rightarrow$ 2019-08-26 \\ 	bc1q0n & \N		& 2 & Back &   2   & 2019-12-09 $\rightarrow$ 2019-12-09 \\ 	bc1qzs & \N		& 2 & Back &  15   & 2019-06-29 $\rightarrow$ 2019-07-21 \\ 	35KxqL & \N		& 3 & Back &   2   & 2019-12-10 $\rightarrow$ 2019-12-10 \\ 	3Dmcs6 & \N		& 3 & Back &   1   & 2019-12-09 $\rightarrow$ 2019-12-09 \\
\hline
\end{tabular}
	\caption{Pony signaling addresses. Exploring from 4 seeds, 8 previously unknown signaling addresses were discovered.}
\label{tab:pony}
\vspace{-0.5cm}
\end{table}

\ignore{
1BkeGq & \cite{pony} & 0 & Seed & 246 & 2019-08-28 & 2021-01-13 \\ 1CeLgF & \cite{pony} & 0 & Seed & 107 & 2019-08-27 & 2019-12-11 \\ 19hi8B & \cite{pony} & 0 & Seed & 352 & 2019-12-10 & 2021-01-13 \\ 1N9ALZ & \cite{ponyPeppermalware}	& 0 & Seed & 19 & 2019-10-03 & 2020-08-06 \\ bc1qwl & \N	& 2 & Forw & 2 & 2019-07-24 & 2019-07-24 \\ 1NL8QT & \N	& 2 & Forw & 3 & 2020-03-03 & 2020-07-22 \\ 1GUghN & \N	& 2 & Back & 6 & 2020-03-03 & 2020-07-11 \\ bc1qh9 & \N & 2 & Back & 28 & 2019-06-09 & 2019-08-26 \\ bc1q0n & \N & 2 & Back &  2 & 2019-12-09 & 2019-12-09 \\ bc1qzs & \N & 2 & Back & 20 & 2019-06-29 & 2019-07-22 \\ bc1q89 & \N & 3 & Back &  7 & 2019-08-09 & 2019-08-15 \\ }

\paragraph{Pony.}
The Pony downloader and information stealer 
started using the Bitcoin blockchain for signaling C\&C servers 
in August 2019~\cite{pony}.
The Pony oracle is detailed in Algorithm~\ref{alg:pony}.
Given an input address, the oracle tries to find pairs of
consecutive transactions that deposit into the address,
whose values encode a public (i.e., not private or reserved) IP address
using a known encoding~\cite{ponyCheckpoint}.
Each transaction in a valid pair encodes two bytes of the IP address. 
The oracle iterates over all the input address transactions (line 4). 
Each deposit transaction 
should have at most three inputs and at most two outputs (line 6).
The oracle temporarily stores the value of the first transaction (line 21)
and validates the transaction values if the second transaction happens 
within an hour of the first one (line 13).
Both transaction values (in satoshis) are passed 
to the \emph{decode} function 
described in~\cite{ponyCheckpoint} (line 14).
If a public IP address is decoded, it is stored (lines 15--16).
The current pair is reset (tx1 set to None)
if a non-public IP address is found, any check fails, or
if a withdrawal transaction is encountered
since the two deposit transactions should be consecutive.
If at least half of all deposit transactions received by the input address 
signal a public IP address, 
the input address is considered a signaling address (line 30).
This last check handles high volume addresses where a pair of 
consecutive deposits may decode into a public IP address by chance.
It also accounts for signaling addresses that were used for 
other purposes before they started signaling.

An exploration is performed starting from four seeds 
obtained from two sources~\cite{pony,ponyPeppermalware}. 
The operation oracle identifies 8 previously unknown signaling addresses. 
The 12 signaling addresses are detailed in Table~\ref{tab:pony}.
Discovering 6 of the 8 new addresses requires at least one backwards 
exploration step, which again highlights the importance of \exploration.
Four of the new addresses are SegWit (starting prefix $bc1$).
They seem to be used for testing before 
switching to the production non-SegWit addresses (starting prefix $1$). 
For example, $bc1q0n$ generates two IP addresses on December 9th, 2019. 
The next day, $19hi8B$ starts signaling IP addresses and 
the last IP signaled by $bc1q0n$ is the first IP signaled by $19hi8B$.
The other SegWit addresses behave similarly.
Using SegWit addresses for testing could be motivated by smaller 
transaction fees.

In addition to the new signaling addresses, backwards exploration 
also finds that the funds used by two signaling transactions 
come from the MINE exchange~\cite{mineExchange}.
This provides a previously unknown attribution point, 
illustrating how backwards exploration can find important relations that 
\tracing cannot.

Encoding the C\&C servers on deposit transactions 
has the downside that analysts can deposit funds into a 
signaling address to hijack the C\&C communication.
On August 14th, 2020, 
Taniguchi et al.~\cite{pony} performed a takeover on 
Pony by signaling a server under their control to $1Bke4Gq$. 
However, they did not sustain the takeover, 
so the attackers regained control two days later.

\begin{table}[t]
\centering
\scriptsize
\begin{tabular}{|l l r l r l|}
\hline
	\textbf{Address} & \textbf{In} & \textbf{Step} & \textbf{Found} & \textbf{FQDNs} & \textbf{Signaling Period} \\
\hline

	15y7ds & \cite{gluptebaSophos}   & 0 & Seed & 8 & 2019-06-19 $\rightarrow$ 2020-05-13 \\
	1CgPCp & \cite{gluptebaSophos}   & 0 & Seed & 6 & 2020-04-08 $\rightarrow$ 2021-10-19 \\
	1CUhaT & \cite{gluptebaTakedown} & 0 & Seed & 6 & 2021-10-13 $\rightarrow$ 2021-12-29 \\
	34Rqyw & \N		 & 1 & Forw & 1 & 2020-01-23 $\rightarrow$ 2020-01-23 \\
	3NhC1b & \N		 & 2 & Forw & 1 & 2020-01-24 $\rightarrow$ 2020-01-24 \\

\hline
\end{tabular}
	\caption{Glupteba signaling addresses. Exploring from 3 seeds, 2 previously unknown signaling addresses were discovered.}
	\vspace{-0.6cm}
\label{tab:glupteba}
\end{table}

\paragraph{Skidmap.}
Skidmap is a crypto mining malware targeting 
Linux hosts~\cite{skidmapTrendMicro}.
On February 2021, Akamai researchers identified that Skidmap
had started using Pony's signaling as a backup C\&C mechanism, and 
mentioned one signaling address~\cite{skidmapAkamai}.
We reuse Pony's oracle (Algorithm~\ref{alg:pony}) for Skidmap.
That seed 
is an online wallet. 
Thus, forward tracing from the seed cannot be performed. 
But, the exploration can proceed backwards from the seed.
In this case the exploration does not find any new signaling address. 
However, in Section~\ref{sec:relations}, 
we show how the backwards exploration identifies
that the funds used in the signaling come from a previous Skidmap 
crypto-jacking campaign and
that operators cashed-out part of the funds in three exchanges that
can serve as attribution points,
again illustrating how \exploration identifies important 
relations \tracing could not.

\paragraph{Glupteba.}
Glupteba is a botnet that has been operating since 
at least 2011~\cite{gluptebaEarly}. 
In September 2019, it was reported to have added Bitcoin as a 
mechanism to distribute backup C\&C domains~\cite{gluptebaTrendMicro}. 
The C\&C signaling leverages Bitcoin transactions with an OP\_RETURN output, 
which stores in the blockchain an AES-GCM-256 encrypted backup C\&C domain. 
The Glupteba oracle, detailed in Algorithm~\ref{alg:glupteba}, 
takes as input the address to validate and two decryption keys
obtained from the malware binaries~\cite{gluptebaSophos}.
It first checks if the given address has a withdrawal transaction (line 2)
with an output (line 3) with the field \textit{data} (line 5),
which indicates an OP\_RETURN slot.
If so, it checks that length of the data field is at least 56
hexadecimal characters (28 bytes).
If so, it attempts to decrypt the data with the input keys,
by using the first 12 bytes as the initialization vector (line 10),
the last 16 bytes as the GCM tag (line 11),
and the rest as the encrypted data (line 12).
If the decryption succeeds (line 14),
the decrypted plain text corresponds to the C\&C domain.

An exploration is performed starting from three seeds
obtained from prior reports~\cite{gluptebaSophos,gluptebaTakedown}.
The operation oracle identifies 2 previously unknown signaling addresses in
the graph.
The 6 signaling addresses are detailed in Table~\ref{tab:glupteba}.
The two discovered addresses seem to have been used for testing. 
They signal the same domain in consecutive days and one day later 
that domain is signaled by address $15y7ds$.
In December 2021, Google performed a take-down on the Glupteba botnet, 
although their note mentions that thanks to the Bitcoin C\&C signaling the 
malware authors are likely to regain control of the 
botnet~\cite{gluptebaTakedown}.

\ignore{
\paragraph{Oracle accuracy.}
We have manually validated that all addresses identified by the oracles 
are indeed TPs, with no FPs, for a precision of 100\%. 
TPs have been validated because they correspond to 
(1) seed addresses known to be signaling, 
(2) new addresses signaling domains/IPs also seen in known 
signaling addresses, and 
(3) producing domains/IPs labeled as malicious in VirusTotal and 
contacted by samples labeled as belonging to the malware family. 
Since we do not have ground truth we cannot guarantee there are no FNs and 
calculate the recall, but all seed addresses from prior works are 
correctly identified, and our manual analysis does not find any 
missing signaling addresses in the graphs. 
Note that we do not claim the operation oracles have perfect accuracy. 
It is possible that if they were applied on a huge number of transactions 
(e.g., on the whole blockchain) FPs would appear. 
However, operation oracles do not affect the relations found as 
we only use them to validate that the graphs include 
previously unknown signaling addresses.
}

In summary, 
the results show \exploration finds new C\&C signaling 
addresses and important relations that \tracing could not. 
In particular, of the 13 new signaling addresses discovered, 
8 are only discovered through backwards exploration, 
and the 4 attribution points found in Skidmap and Pony are also only 
discovered through backwards exploration.

\ignore{
\paragraph{BitMEX exchange}
BitMEX~\cite{bitmex} is a cryptocurrency exchange and derivative trading platform
with offices worldwide, active since 2014.
While producing tags, we observed a set of addresses~\cite{bitinfocharts}
beginning with the prefix \textit{3BMEX} associated to BitMEX~\cite{bitmex-wallet}
that rarely produce multi-input transactions.
Clusters associated to these addresses are commonly singletons,
so tagging individual addresses of BitMEX
is not helpful to detect other addresses
sharing the same owner, as is common with other exchanges
having large multi-input clusters.

As described in its website, BitMEX uses multisig addresses
to kept customer funds.
By analyzing withdrawal transactions from several
BitMEX multisig addresses, we found that
all of them are scripthash addresses of type multisig 3-of-4,
so three addresses from a total of four are required to
produce valid transactions;
Their wrapped addresses are always of type pubkey;
From each set of four pubkey addresses wrapped in a BitMEX multisig,
there is a common subset of three pubkey addresses,
namely 158Yaf, 13xiA2, and 1EczoQ, listed in
Table~\ref{tab:mapping} of the appendix.

Using this information, we build an oracle to identify BitMEX addresses
that signed at least one withdrawal transaction,
that should exist to be able to identify the addresses
wrapped in the multisig scripthash address.
The proposed approach is consistent with
addresses produced by BitMEX having withdrawal transactions,
for instance its cold-wallet~\cite{bitmex-wallet}.
}

\begin{figure}[tb]
  \centering
  \includegraphics[width=55mm]{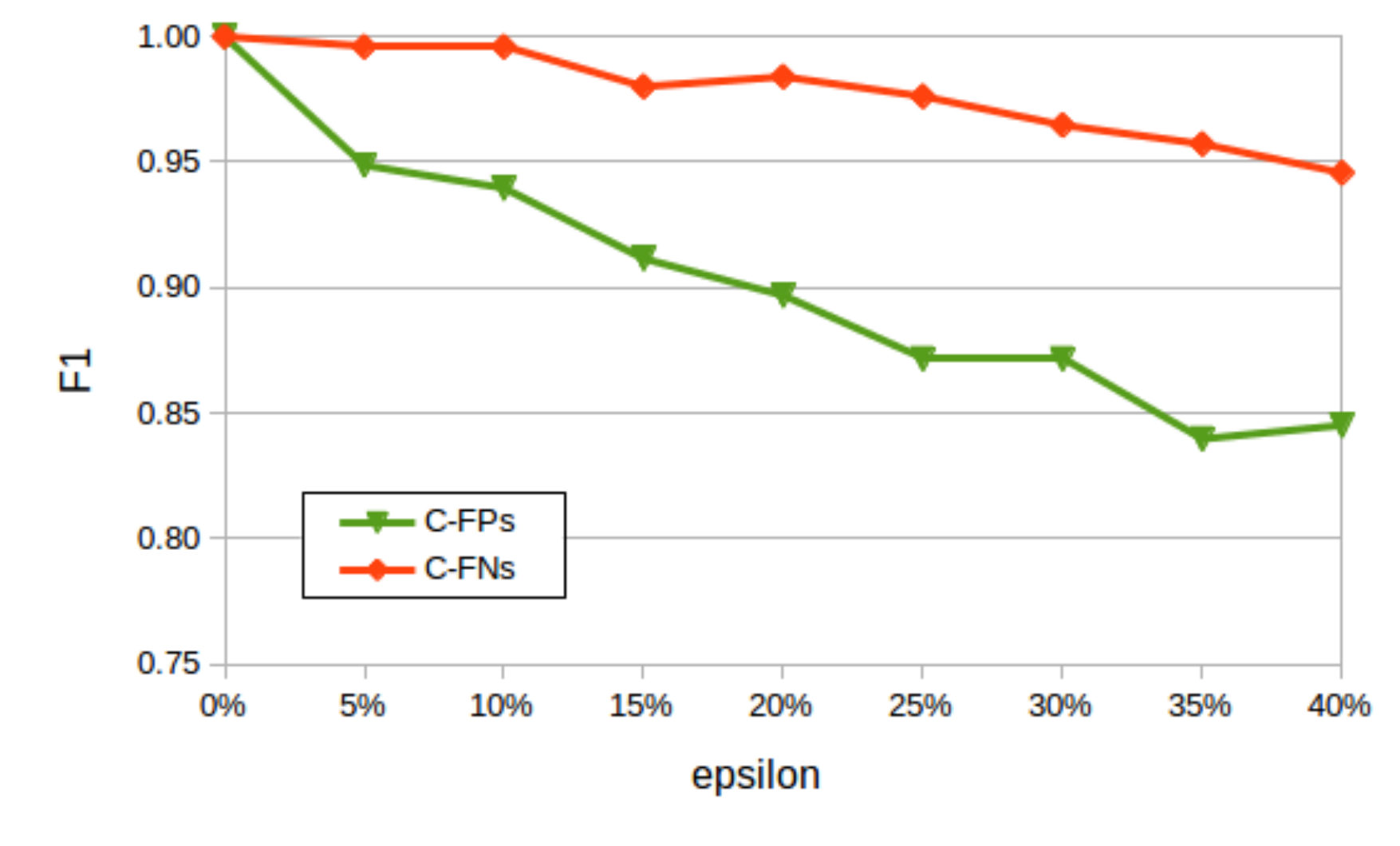}
\vspace{-0.3cm}
	\caption{Relation detection impact 
when introducing exchange classifier errors. 
Introducing C-FPs produces R-FNs (green line).
Introducing C-FNs produce R-FPs (red line).}
  \label{fig:epsilon}
\end{figure}

\subsection{Exchange Classifier Impact}
\label{sec:eval_classifier}

This section evaluates the impact of the exchange classifier 
in the \exploration. 
First, we measure how exchange classification
errors impact the identification of end-to-end relations.
Then, we perform an ablation study to measure the performance impact
if the exchange classifier is not used.
Both experiments are run on the 23 explorations with at least one 
seed that is not an online wallet. 

\paragraph{Accuracy impact.}
A false positive by the exchange classifier (C-FP) might 
make the exploration discard valid paths and thus miss true relations,
introducing end-to-end false negatives (R-FN).
On the other hand, an exchange classifier false negative (C-FN) 
leads to larger graphs that 
may contain false relations introducing end-to-end false positives (R-FP).
To measure the impact of exchange classification 
errors on the identification of end-to-end relations,
we run the \exploration multiple times on each family. 
Each exploration introduces errors to the exchange classifier results 
with an $\epsilon$ probability, and compares the relations 
found to the family's baseline in Table~\ref{tab:graphs}. 
An R-FN is the absence of a relation 
in the baseline, 
while an R-FP is a new relation not present in the baseline.

For ease of interpretation, we perform two experiments: 
introducing only C-FPs and 
introducing only C-FNs.
For example, when introducing C-FPs, 
if the classifier says an address is not an exchange, 
with $\epsilon$ probability the decision is flipped to mark it as an exchange.
We increase $\epsilon$ between 5\% and 40\% in steps of 5\%.
Figure~\ref{fig:epsilon} shows, for each experiment,
the F1 score of the relation detection as a function of $\epsilon$.
The results show that the introduction of C-FPs
creates a more pronounced decrease in the relation detection results.
Mis-classifying a non-exchange address as an exchange (C-FP) 
stops the exploration too early, which causes relations to be missed (R-FNs).
On the other hand, C-FNs introduce additional addresses in the graph, 
but since many of those new addresses are untagged, the number of R-FPs 
introduced is lower in comparison, and thus the F1 score reduces more slowly.
These results indicate that favoring a high precision of 
the exchange classifier is beneficial to the overall relation detection. 
This is already true for our classifier as 
Table~\ref{table:classification} shows that its precision is 
higher than its recall.

\begin{table}[t]
\centering
\scriptsize
\begin{tabular}{|l|rr|rr|}
\cline{2-5}
\multicolumn{1}{c|}{} & \textbf{Runtime} & \textbf{Timeout} & \textbf{Addresses} & \textbf{TXES} \\
\hline
With exchange classifier	&  499'37" & 0 &  52,308 & 145,410 \\ Without exchange classifier & 2598'34" & 7 & 243,019 & 787,397 \\ \hline
Increase w/o exchange classifier & 5.2x & 7 & 4.6x & 5.4x \\
\hline
\end{tabular}
\caption{Total runtime and graph size, 
across the 23 families where at least one seed is not an online wallet,
when running \exploration with and without the exchange classifier.
To handle explosion due to not using the classifier, 
7 explorations were stopped when their graph reached 20K addresses.
}
\label{tab:ablation}
\vspace{-0.6cm}
\end{table}

\paragraph{Ablation study.}
This experiment runs the \exploration on each family 
without using the exchange classifier. 
This is equivalent to the classifier introducing 100\% C-FNs.
The removal of the classifier causes some explorations to explode,
i.e., they fail to finish within 12 hours.
To be able to complete the experiment in reasonable time, 
we stop an exploration when its graph reaches 20k addresses.
Table~\ref{tab:ablation} summarizes the impact of removing the exchange 
classifier in terms of runtime and graph size.
When the exchange classifier is not used, 
the graphs are on average 4.6 times larger 
and the runtime 5.2 times larger.
In 7 of the 23 explorations, the 20K address limit is reached. 
Thus, the performance impact for those 7 explorations 
would be larger in reality.
These results show how the exchange classifier is crucial
to avoid explosion by limiting the runtime and graph size.
The removal of the exchange classifier also causes the R-FPs to increase,
as illustrated in the previous experiment, 
but the 20K address limit prevents us from precisely measuring the F1 decrease.

\begin{table*}
\centering
\scriptsize
\begin{tabular}{|l|r|lp{0.75\linewidth}|}
\hline
\textbf{Class} & \textbf{Operation} & \multicolumn{2}{l|}{\textbf{Relations Found}} \\
\hline
  \multirow{3}{*}{\rotatebox{90}{\makecell{C\&C}}}
 & cerber          & 1 & malware:cerber:ransomnote \\
 & glupteba        & 0 & - \\
 & pony            & 1 & exchange:mineexchange \\
 & skidmap         & 5 & cryptojacking:sysupdate, exchange:bitfinex, exchange:coinbase, exchange:coincola*, exchange:huobi \\
\hline
  \multirow{17}{*}{\rotatebox{90}{Clipper}} 
 &aggah            & 1 & exchange:binance \\
 &androidclipper   & 1 & exchange:luno.com* \\
 &azorult          & 1 & exchange:tradeogre* \\
 &bitcoingrabber   & 0 & - \\
 &casbaneiro	   & 4 & exchange:bleutrade.com, exchange:localbitcoins, exchange:exmo, payment:coinpayments \\
 &clipbanker       & 1 & exchange:coinbase* \\
 &clipsa           & 6 & exchange:binance, exchange:bitfinex, exchange:changenow, exchange:localbitcoins, mixer:wasabi, payment:coinpayments \\
 &cliptomaner      & 1 & exchange:tradeogre* \\
&cryptoshuffler   & 11 & exchange:bittrex, exchange:btc-e.com, exchange:btctrade.com, exchange:cryptopay.me, exchange:cubits.com, exchange:exmo, exchange:poloniex, exchange:spectrocoin, mining:btcc-pool, payment:bitpay.com, payment:coinpayments \\
 &masad           & 11 & exchange:any-cash, exchange:binance, exchange:bitzlato, exchange:coinbase, exchange:cryptonator, exchange:localbitcoins, exchange:whitebit, exchange:z-exchange.ru, gambling:1xbit, payment:coinpayments, payment:payeer \\
 &mekotio          & 6 & clipper:mekotion40*, exchange:hitbtc, exchange:mercadobitcoin, miscabuse:mrpr0gr4mmer, mixer:wasabi, payment:coinpayments \\
 &mekotion40       & 6 & clipper:mekotio*, exchange:dsx, exchange:mercadobitcoin, gambling:foxbit, gambling:yolo-group, mixer:wasabi \\
 &mispadu          & 1 & exchange:mercadobitcoin* \\
 &mrpr0gr4mmer     & 2 & exchange:binance, payment:coinpayments \\
 &n40              & 1 & exchange:mercadobitcoin* \\
 &phorpiex-tldr    & 3 & exchange:cryptonator*, gambling:duckdice.io, miscabuse:saveyourself \\
 &phorpiex-trik    & 5 & exchange:bitcoin.de, exchange:localbitcoins, exchange:cryptonator*, gambling:rollin.io, payment:webmoney \\
 &predatorthethief & 8 & clipper:masad, exchange:binance, exchange:bitzlato, exchange:btcbank, exchange:cryptonator, exchange:totalcoin, exchange:whitebit, payment:coinpayments \\  &protonbot        & 2 & exchange:cryptonator*, exchange:hitbtc* \\
\hline
  \multirow{15}{*}{\rotatebox{90}{Ransomware}} 
 &cryptotorlocker2015& 11 & exchange:bitcoin.de, exchange:cex.io, gambling:betcoin-dice, gambling:fortunejack.com, gambling:luckybit, gambling:satoshibones, gambling:satoshidice.com, mining:cointerra, mining:deepbit, mining:eobot, tormarket:silkroad2market \\
 &cryptxxx         & 13 & exchange:bitbay.net, exchange:bitcoin.de, exchange:bitfinex, exchange:bitflyer, exchange:btc-e.com, exchange:coinbase, exchange:coinmate.io, exchange:coins.co.th, exchange:huobi, exchange:localbitcoins, exchange:okcoin.com, exchange:poloniex, exchange:xapo \\
 &dmalocker        &  8 & exchange:btc-e.com, exchange:localbitcoins, exchange:okex, exchange:poloniex, gambling:cloudbet.com, malware:slave, payment:bitpay.com, payment:coinpayments \\
 &globeimposter    &  9 & exchange:bitstamp, exchange:bittrex, exchange:btc-e.com, exchange:cex.io, exchange:localbitcoins, payment:bitpay.com, payment:easywallet, payment:webmoney, tormarket:evolutionmarket \\  &locky            & 15 & exchange:bitfinex, exchange:bitstamp, exchange:btc-e.com, exchange:coinbase, exchange:localbitcoins, exchange:matbea.com, exchange:poloniex, exchange:coins.co.th, exchange:exmo, exchange:xzzx.biz, mixer:bitcoinfog, payment:bitpay.com, payment:coinpayments, payment:webmoney, tormarket:nucleusmarket \\
 &samsam           &  6 & exchange:bitcoin.de, exchange:cubits.com, exchange:localbitcoins, exchange:poloniex, exchange:yobit.net, tormarket:alphabaymarket \\
 &wannacry	   &  2 & exchange:changelly, exchange:shapeshift \\
\hline
\end{tabular}
\caption{Summary of relationships found. 
	Tags with an asterisk at the end were found by multi-input clustering on the seeds, the rest were found by the exploration.}
\label{tab:relations}
\vspace{-0.7cm}
\end{table*}

\section{Relationships}
\label{sec:relations}

Tagged addresses in a \graph may capture
relations between the campaign under analysis and 
external services (e.g., exchanges, mixers, gambling, markets) and
other malicious campaigns.
Those relations are identified by the graph analysis module
as paths that detail the specific chain of 
transactions capturing money flow 
from the campaign under analysis to an external entity, or vice-versa.
Such paths constitute evidence that can be shared with third parties
to allow independent validation of the relations.

Table~\ref{tab:relations} summarizes the relations found.
In 93\% (28/\numoperations) of the explorations at least one relation 
is identified,
which we believe reflects that our tag database provides good coverage.
Tags may have been assigned to addresses found by MI clustering 
on the seeds (9.2\%)
(marked with asterisk at the end of the tag) or 
during the exploration (90.8\%). 
The large percentage difference highlights the need to perform 
transaction tracing (beyond MI clustering) to discover relations.
The most common relations are with exchanges
used by the operators for seeding C\&C signaling and cashing out profits.
We observe 44 exchanges, the most popular one being
\tag{localbitcoins} used by 9 families, followed by 
\tag{coinbase} (5), \tag{poloniex} (5), \tag{cryptonator} (5), \tag{btc-e} (5), and
\tag{binance} (5).
Ransomware families
use a larger variety of exchanges compared to clippers, 
e.g., the \tag{cryptxxx} ransomware uses 13 exchanges.
We also observe 5 payment processors, 
also used for cashing out: 
\tag{coinpayments} (9), \tag{bitpay} (4), 
\tag{webmoney} (3), \tag{easywallet} (1), and \tag{payeer} (1). 
These services used for cashing out 
include some that are not amongst the most popular
ones~\cite{forbesExchanges2021}, 
which may be due to relaxed KYC measures.
For example, \tag{localbitcoins} is a peer-to-peer exchange 
that allows users to transact between them over-the-counter 
in a fairly anonymous manner.
\tag{Coinpayments} is a payment gateway
that merchants can use to receive cryptocurrency
with just an email and password.
Both services did not establish KYC measures until
mid-2019~\cite{coinpaymentsKYC,localbitcoinsKYC}.
Furthermore, the owner of the \tag{btc-e} exchange
was arrested in July 2017 for money laundering 
with the indictment mentioning the lack of 
KYC measures~\cite{btceTakedown}.

Other common relations are gambling sites
and mixers,
likely used for money laundering.
Six families are observed transacting with 11 gambling sites
(e.g., \tag{satoshidice}) 
with the \tag{cryptotorlocker2015} ransomware using a record 5 gambling sites.
Gambling sites are also possible attribution points since
they may be subjected to legislation mandating KYC practices 
such as the 
Licence Conditions and Codes of Practice (LCCP) in the UK~\cite{gamblingUkKyc}.
Two mixers appear across 4 families:
\tag{wasabi} (3) and \tag{bitcoinfog} (1).
Wasabi~\cite{wasabi} is an open-source Bitcoin wallet 
implementing CoinJoin shuffling over Tor for anonymity.
The operator of \tag{bitcoinfog} 
was arrested on April 2021 on counts of 
unlicensed money transmission and money laundering~\cite{bitcoinfogTakedown}.
Other relations found include 
three malware families using profits for buying products from 
four Tor hidden markets 
(\tag{alphabaymarket}, \tag{evolutionmarket}, \tag{nucleusmarket}, 
\tag{silkroad2market})
and two families transacting with four mining pools
(\tag{btcc-pool}, \tag{cointerra}, \tag{deepbit}, \tag{eobot}). 
Finally, several relations between malicious campaigns are found, 
as detailed next.

\begin{figure}[t]
  \centering
  \includegraphics[width=\columnwidth]{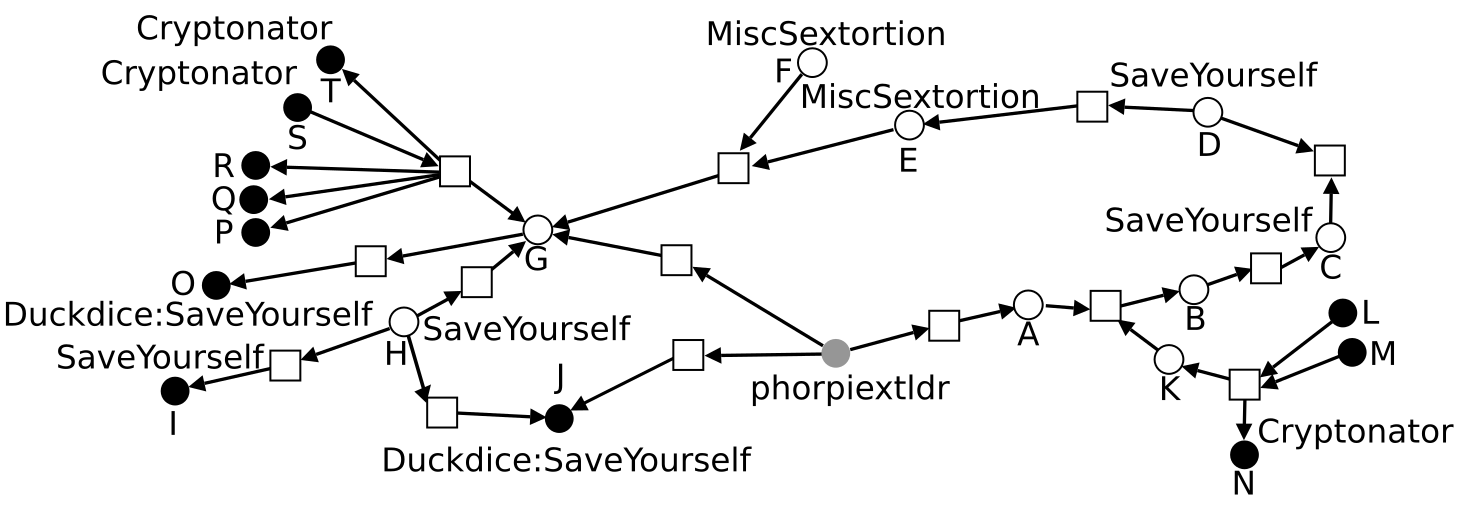}
  \vspace{-0.8cm}
\caption{Phorpiex-tldr -- SaveYourself relations.}
  \label{fig:phorpiextldr}
  \vspace{-0.3cm}
\end{figure}

\paragraph{Phorpiex-tldr and SaveYourself.}
The exploration from \tag{phorpiex-tldr} 
reveals multiple connections with the \tag{saveyourself} sextortion campaign.
Figure~\ref{fig:phorpiextldr} shows the paths capturing these relations, 
other paths are excluded for readability.
There are six \tag{saveyourself} tagged addresses in the graph, 
corresponding to addresses reported by Bitcoin Abuse users 
as appearing in sextortion emails with a
common SaveYourself pattern in the subject and content. 
The figure shows how the clipper's profits flow towards multiple 
\tag{saveyourself} addresses through different paths, e.g., 
to tagged address C from the seed (in gray) through addresses A and B,
to tagged address J from the seed,
and through collector node G to tagged address O.
Addresses O and J correspond to online wallets in the \tag{duckdice.io}
gambling service used by SaveYourself operators for betting.
Some connections are found through \tracing, 
but others such as the path going from the seed to address D through 
A, B, and C
or the path going from the seed through G and H to reach node I
require exploring backwards.
This illustrates the benefits of \exploration.
An external report mentions that the Phorpiex malware 
sends sextortion emails~\cite{saveYourself}.
Checking the Bitcoin addresses in that report on Bitcoin Abuse confirms 
the link between Phorpiex and the SaveYourself campaign.  
Furthermore, the multiple connections between both campaigns 
and how the clipper profits flow into the sextortion addresses 
makes us conclude that the sextortion campaign is run by the malware operators,
as opposed to the Phorpiex operators renting their botnet 
for sending third-party spam.

\paragraph{MrPr0gr4mmer clipper and scam.}
The MrPr0gr4mmer exploration, summarized in Figure ~\ref{fig:mekotiomr},
starts from two seeds (A, B) in the same MI cluster, 
each with multiple clipper reports in Bitcoin Abuse.
Address F is reported in sextortion scam emails~\cite{mrpr0gr4mmerScam}.
The sender of the scam emails is ``Mr Robot''.
Manual searches uncover that 
seed A appears in the @MrPr0gr4mmer Telegram channel, 
where software to send spam through the
telegram API~\cite{mrpr0gr4mmer} is advertised.
The contact address starts again with the ``Mr.'' prefix. 
Seed B appears as donation address for 
a user of
the \url{cracked.to} forum and 
backwards tracing finds that collector E receives deposits 
from the \tag{paxful} exchange.
We conclude that 
the MrPr0gr4mmer operator
could be attributed through the \tag{paxful} exchange and the 
\url{cracked.to} forum and
is involved in multiple malicious activities 
including selling spam tools, sending scam emails, and 
operating a clipper. 

\begin{figure}[t]
  \centering
  \includegraphics[width=0.9\columnwidth]{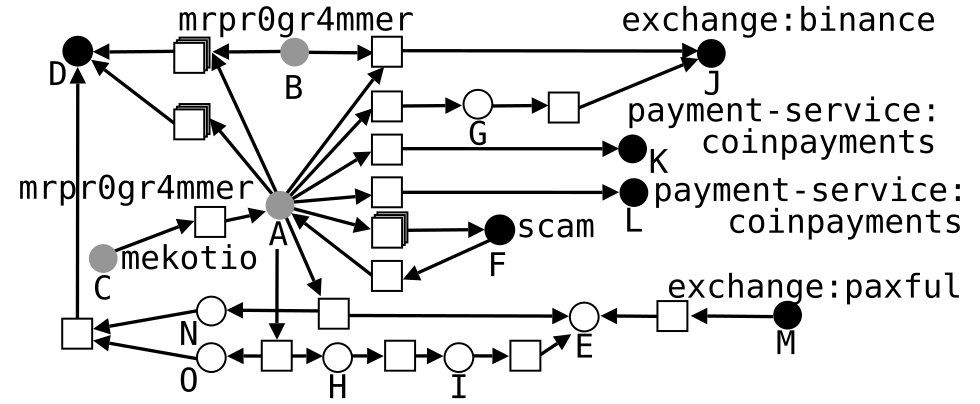}
  \caption{Mekotio-MrPr0gr4mmer relations.}
  \label{fig:mekotiomr}
\vspace{-0.4cm}
\end{figure}

\paragraph{Mekotio, MekotioN40 and MrPr0gr4mmer.}
A seed from \tag{mekotio} and a seed from \tag{mekotion40} 
appear as inputs to the same transaction indicating that 
both families are indeed the same. 
While the relationship was already hinted by the names assigned in the 
two reports from where we obtained the seeds, 
those reports do not explicitly mention a connection. 
In addition, both graphs show cash outs to the 
\tag{mercadobitcoin.br} exchange and transactions to the 
\tag{wasabi} mixer in the non-shared components, 
further reinforcing that both campaigns have the same operators. 
Another relation in the \tag{mekotio} graph is a payment from 
a \tag{mekotio} seed to the seed of \tag{mrpr0gr4mmer} used for 
selling the Telegram spam tool, illustrated in Figure~\ref{fig:mekotiomr}.
Thus, \tag{mekotio} operators are using the obtained profits to pay for a 
spam tool possibly used as part of their malware distribution.

\paragraph{PredatorTheThief and Masad.}
The \tag{predatorthethief} clipper seed makes a
transaction to an address that a report 
identifies as belonging to the 
Masad Clipper and Stealer malware~\cite{masadReport}. 
Masad is offered as a malware kit in forums.
It has a free version and a full version that costs \$85.
The transaction on June 2020 is for 0.00910768 BTC, 
roughly \$85.31 USD using the conversion at that time. 
We believe this relation captures how \tag{predatorthethief} operators 
bought a license of \tag{masad}, possibly to upgrade their own malware.

\paragraph{DMALocker and Slave.}
The \tag{dmalocker} ransomware graph contains a transaction 
to an address tagged as belonging to the 
Slave banking trojan~\cite{slaveReport}.
Slave receives 2.3 BTC from DMALocker, more than 1.3K USD at that time. 
Unfortunately, in this case we are not able to conclude what type of 
relation the transaction captures.

\paragraph{Skidmap and Sysupdate.}
The \tag{skidmap} cryptominer backwards exploration 
(\tracing is not possible as the seed is an online wallet) 
identifies a \tag{cryptojacking:sysupdate} address
that sends funds to the seed six times.
This address belongs to a mining pool used by a Linux 
cryptomining malware to receive the shares produced by the botnet's 
mining activity~\cite{sysupdate}.
We believe this malware is an older \tag{skidmap} version
(prior to using Bitcoin for C\&C) 
and that the relation shows that funds from Skidmap's cryptomining were used 
to seed the new Bitcoin C\&C signaling.
The Sysupdate address cashes out in three 
exchanges (\tag{coinbase}, \tag{huobi}, \tag{bitmex}) 
that can be used as attribution points and are only found exploring backwards.

\paragraph{CryptoTorLocker2015 and Scam Campaigns.}
The MI cluster of the \tag{cryptotorlocker2015} ransomware seeds 
contains 10 addresses beginning with short English words
involved in multiple scam campaigns in 2014-2015. 
Four of those imitate satoshidice gambling addresses beginning with ``1DiCE'',
another four 
(e.g., 1BonUS, 1sYSTEM) were used in scams that promise to double the 
BTCs sent~\cite{bonusScam},
and two addresses beginning with ``1bet'' and ``1shop''
were used in campaigns claiming to have hacked the 
bitcointalk forum~\cite{betScam}.
Those scams are likely run by the ransomware operators.

\subsection{Limited Ground Truth Validation}
\label{sec:bitfog}

Ground truth is not available for cybercrime operations and 
can only rarely be obtained, e.g., through leaks~\cite{pharmaleaks,contiLeaks}. 
Still, we tried hard to find some ground truth, even if limited, 
encountering two major challenges. 
First, prior analysis focuses on forward-only explorations 
and thus does not mention relations only reachable by exploring backwards.
Second, analysis reports rarely exhaustively list relations, 
mentioning only those of interest to the report. 
We found only one report that partially addresses those challenges, 
which we use as a limited ground truth.  
It corresponds to the affidavit 
filed in April 2021 by the US Internal Revenue Service
to justify the takedown of 
the \tag{bitcoinfog} mixer~\cite{bitcoinfogAffidavit}.
The affidavit mentions \tag{bitcoinfog} received deposits 
from 35 darknet markets, 
but only names the five largest: 
\tag{agora}, 
\tag{alphabay},
\tag{evolution},
\tag{silkroad}, and 
\tag{silkroad2}.

Our database has one address tagged as \tag{bitcoinfog} belonging to a 
MI cluster with 244K addresses. 
Forward exploration from the mixer addresses does not work as 
the mixer's goal is to hamper \tracing.
Thus, we perform a backwards-only exploration using as seed all addresses 
in the \tag{bitcoinfog} MI cluster. 
The exploration finds a wealth of relations 
including all 15 darknet markets in our database.
Those 15 markets include the five named in the affidavit
and another 10:
\tag{abraxas}, 
\tag{babylon},
\tag{blackbank},
\tag{bluesky}, 
\tag{cannabisroad},
\tag{doctord}, 
\tag{middleearth}, 
\tag{nucleus},
\tag{pandora}, and
\tag{sheep}.

Since all 15 tagged darknet markets are found after one backwards 
step, we stop the exploration at that point.
In addition to the darknet markets, 
the backwards exploration identifies relations with 130 services and 
15 malicious operations not mentioned in the affidavit: 
4 ransomware (\tag{cryptolocker}, \tag{locky}, \tag{xlocker}, \tag{xtplocker}),
3 scams (\tag{mintpal}, \tag{mcxnow}, \tag{bitoomba}), 
3 thefts (\tag{binance-hack}, \tag{bips-hack}, \tag{flexcoin-hack}).
3 mixers (\tag{bitlaunder}, \tag{wasabi}, and an unknown mixer),
a malware (\tag{slave}), and
a ponzi scheme (\tag{nanoindustryinv}).

This experiment shows backwards exploration finds correct relations, 
although it may miss relations for which no tags are available. 
Unfortunately, we cannot evaluate false positives as the affidavit 
does not list all markets, and other services and malicious operations.
\section{Discussion}
\label{sec:discussion}

\paragraph{Backwards exploration.}
Our results show that \exploration can discover important relations 
that \tracing cannot.
One limitation is that backwards exploration is 
not suitable for every transaction. 
However, we show how it is possible to identify the problematic cases. 
For example, we provide a simple heuristic to identify dust attacks 
that may deposit small amounts to many addresses.
We also provide an option to disable backwards exploration on seeds, 
which requires an easy decision by the analyst: 
use it when seeds may get deposits from 
other addresses belonging to the same campaign (e.g., C\&C signaling) 
whereas avoid it when deposits come from victims 
(e.g., clippers, ransomware). 
Furthermore, even if a transaction was included that it should not, 
the transaction and its dependencies can later be pruned from the graph.

\paragraph{Exchange defenses.}
90\% of malware families analyzed had 
at least one relation with an exchange. 
This includes 10 malware families where addresses hardcoded in the malware
correspond to online wallets in exchanges, 
These exchanges are either unaware about the abusive nature of those 
addresses, or they simply do not care.
This creates an opportunity for intervention where 
public malware analysis reports could be automatically analyzed 
to extract tagged Bitcoin addresses
(e.g., using indicator extraction approaches~\cite{liao2016acing,chainsmith}). 
The identified addresses could then be automatically explored, and 
those identified as online wallets or receiving cash-outs from the 
malware activities could be reported to the exchanges. 
This would prevent exchanges from claiming they were unaware of the abuse, 
and allow evaluating the reaction of exchanges to abuse reports.

\paragraph{Address classification.}
Tagging Bitcoin addresses 
is fundamental to identify relations.
A well-known tagging limitation is low coverage, as only a very small 
percentage of addresses is tagged.
Furthermore, tagging is often delegated to commercial services 
(e.g.,~\cite{chainalysis,elliptic}), 
which use proprietary approaches to build their tag databases 
and do not disclose how they handle challenges such as 
double-ownership and potentially conflicting tags. 
We believe open research on building large, accurate, tag databases 
is fundamental to allow tagging to sustain challenges likely to 
appear in court cases.
Furthermore, we have shown how to build classifiers to identify 
certain types of addresses such as those belonging to exchanges and 
those used by malware for C\&C signaling. 
A possible avenue of future work is to build classifiers for many 
other address types. 

\paragraph{Evasion.}
An attacker could evade our approach by making its addresses 
look uninteresting, e.g., like exchanges.
However, feature analysis shows that the high volume of 
transactions is a key feature of the exchange classifier, 
so impersonating an exchange would be expensive. 
Alternatively, an attacker could pollute our tag database 
so that some of its addresses are considered benign services.
Evasion is also possible by using mixers since they prevent forward exploration.
However, we have shown that backwards exploration reveals their clients.
Yet another evasion would pollute the graphs with unrelated transactions, 
e.g., by using sophisticated dusting techniques that our approach 
may fail to filter. 
This would make the graphs less readable, but would not prevent automated 
extraction of relations.

\paragraph{Other blockchains.}
Our approach generalizes to other cryptocurrencies following  
Bitcoin's model. 
BlockSci already supports variants of Bitcoin:
Bitcoin Cash, Dash, Litecoin, Namecoin, and ZCash. 
Supporting them would require extending our tag database 
with addresses for those blockchains and to retrain the exchange classifier.

\vspace{-0.3cm}
\section{Conclusion}
\label{sec:conclusion}

Our study of the financial relationships of \numoperations malware families 
sheds light on cybercrime's financial inner workings.
Our novel \exploration identifies important relations missed by 
\tracing such as 
previously unknown C\&C signaling addresses, 
a wealth of relations to external services and other cybercrime operations, 
and new attribution points.
The results of this work can benefit the community 
in two main ways: (i) the open source tool and tag dataset we will provide 
may serve as basis for future works; and (ii)  
the detailed graphs and relations our approach produces  
could be used as evidence to warrant interventions.
 
\section*{Acknowledgments}
This work has been partially supported by the Madrid regional government 
as part of the program S2018/TCS-4339 (BLOQUES-CM), 
co-funded by EIE Funds of the European Union.
This work has been partially supported by the SCUM Project 
(RTI2018-102043-B-I00) MCIN/AEI/10.13039/501100011033/ERDF  
and by grant PRE2019-088472/MCIN/AEI/10.13039/501100011033, 
ESF Investing in your future.
This work has been partially funded by grant
IJC2020-043391-I/MCIN/AEI/10.13039/501100011033 and European Union NextGenerationEU/PRTR.
This work has been partially funded by the project HACRYPT (N00014-19-1-2292).
Any opinions, findings, and conclusions or recommendations in 
this material are those of the authors or originators, and 
do not necessarily reflect the views of the sponsors.

{
\bibliographystyle{ACM-Reference-Format}
\bibliography{bibliography/paper}


\begin{thebibliography}{71}


\ifx \showCODEN    \undefined \def \showCODEN     #1{\unskip}     \fi
\ifx \showDOI      \undefined \def \showDOI       #1{#1}\fi
\ifx \showISBNx    \undefined \def \showISBNx     #1{\unskip}     \fi
\ifx \showISBNxiii \undefined \def \showISBNxiii  #1{\unskip}     \fi
\ifx \showISSN     \undefined \def \showISSN      #1{\unskip}     \fi
\ifx \showLCCN     \undefined \def \showLCCN      #1{\unskip}     \fi
\ifx \shownote     \undefined \def \shownote      #1{#1}          \fi
\ifx \showarticletitle \undefined \def \showarticletitle #1{#1}   \fi
\ifx \showURL      \undefined \def \showURL       {\relax}        \fi
\providecommand\bibfield[2]{#2}
\providecommand\bibinfo[2]{#2}
\providecommand\natexlab[1]{#1}
\providecommand\showeprint[2][]{arXiv:#2}

\bibitem[bet(2014)]%
        {betScam}
 \bibinfo{year}{2014}\natexlab{}.
\newblock \bibinfo{title}{{BTC-E and Bitcointalk.org Hacked?}}
\newblock
\newblock
\newblock
\shownote{\url{https://www.reddit.com/r/Bitcoin/comments/2mc8nc}}.


\bibitem[bon(2014)]%
        {bonusScam}
 \bibinfo{year}{2014}\natexlab{}.
\newblock \bibinfo{title}{{Warning: New email scam people are apparently
  falling for - "secret doubling address".}}
\newblock
\newblock
\newblock
\shownote{\url{https://www.reddit.com/r/Bitcoin/comments/2fq3uc}}.


\bibitem[cer(2017)]%
        {cerberTakedown}
 \bibinfo{year}{2017}\natexlab{}.
\newblock \bibinfo{title}{European police take down criminals behind two big
  ransomware strains}.
\newblock
\newblock
\newblock
\shownote{\url{https://www.cyberscoop.com/ctb-locker-cerber-ransomware-arrests-europol-mcafee/}}.


\bibitem[pon(2019)]%
        {ponyPeppermalware}
 \bibinfo{year}{2019}\natexlab{}.
\newblock \bibinfo{title}{Brief analysis of Redaman Banking Malware (v0.6.0.2)
  Sample}.
\newblock
\newblock
\newblock
\shownote{\url{http://www.peppermalware.com/2019/11/brief-analysis-of-redaman-banking.html}}.


\bibitem[coi(2019)]%
        {coinpaymentsKYC}
 \bibinfo{year}{2019}\natexlab{}.
\newblock \bibinfo{title}{{Coinpayments - Service Restriction Notice}}.
\newblock
\newblock
\newblock
\shownote{\url{https://bitcointalk.org/index.php?topic=5156686}}.


\bibitem[sav(2019)]%
        {saveYourself}
 \bibinfo{year}{2019}\natexlab{}.
\newblock \bibinfo{title}{In the footsteps of a sextortion campaign}.
\newblock
\newblock
\newblock
\shownote{\url{https://research.checkpoint.com/2019/in-the-footsteps-of-a-sextortion-campaign/}}.


\bibitem[sys(2019)]%
        {sysupdate}
 \bibinfo{year}{2019}\natexlab{}.
\newblock \bibinfo{title}{{Linux mining virus-sysupdate}}.
\newblock
\newblock
\newblock
\shownote{\url{https://programmerclick.com/article/47131149819/}}.


\bibitem[loc(2019)]%
        {localbitcoinsKYC}
 \bibinfo{year}{2019}\natexlab{}.
\newblock \bibinfo{title}{{LocalBitcoins Statement on the Coming AML
  regulations and Compliance}}.
\newblock
\newblock
\newblock
\shownote{\url{https://localbitcoins.com/blog/aml-regulations-compliance}}.


\bibitem[mrp(2020a)]%
        {mrpr0gr4mmerScam}
 \bibinfo{year}{2020}\natexlab{a}.
\newblock \bibinfo{title}{{Bitcoin Porn Scam and Sextortion}}.
\newblock
\newblock
\newblock
\shownote{\url{https://www.onlinethreatalerts.com/article/2020/4/9/bitcoin-porn-scam-and-sextortion/}}.


\bibitem[mrp(2020b)]%
        {mrpr0gr4mmer}
 \bibinfo{year}{2020}\natexlab{b}.
\newblock \bibinfo{title}{{pr0fessionalcrackers telegram channel}}.
\newblock
\newblock
\newblock
\shownote{\url{https://ru.tgchannels.org/channel/MrPr0gr4mmer}}.


\bibitem[bit(2022a)]%
        {bitcoinAbuse}
 \bibinfo{year}{2022}\natexlab{a}.
\newblock \bibinfo{title}{Bitcoin Abuse}.
\newblock
\newblock
\newblock
\shownote{\url{https://www.bitcoinabuse.com}}.


\bibitem[bit(2022b)]%
        {bitcoinExplorer}
 \bibinfo{year}{2022}\natexlab{b}.
\newblock \bibinfo{title}{Bitcoin Explorer}.
\newblock
\newblock
\newblock
\shownote{\url{https://www.blockchain.com/explorer}}.


\bibitem[bit(2022c)]%
        {bitcoin-otc}
 \bibinfo{year}{2022}\natexlab{c}.
\newblock \bibinfo{title}{Bitcoin OTC}.
\newblock
\newblock
\newblock
\shownote{\url{https://bitcoin-otc.com}}.


\bibitem[bit(2022d)]%
        {bitcoinTalk}
 \bibinfo{year}{2022}\natexlab{d}.
\newblock \bibinfo{title}{Bitcoin Talk Forum}.
\newblock
\newblock
\newblock
\shownote{\url{https://bitcointalk.org}}.


\bibitem[cha(2022)]%
        {chainalysis}
 \bibinfo{year}{2022}\natexlab{}.
\newblock \bibinfo{title}{Chainalysis}.
\newblock
\newblock
\newblock
\shownote{\url{https://www.chainalysis.com/}}.


\bibitem[ell(2022)]%
        {elliptic}
 \bibinfo{year}{2022}\natexlab{}.
\newblock \bibinfo{title}{Elliptic}.
\newblock
\newblock
\newblock
\shownote{\url{https://www.elliptic.co/}}.


\bibitem[min(2022)]%
        {mineExchange}
 \bibinfo{year}{2022}\natexlab{}.
\newblock \bibinfo{title}{{MINE.exchange}}.
\newblock
\newblock
\newblock
\shownote{\url{https://mine.exchange}}.


\bibitem[sat(2022)]%
        {satoshiDice}
 \bibinfo{year}{2022}\natexlab{}.
\newblock \bibinfo{title}{Satoshi DICE}.
\newblock
\newblock
\newblock
\shownote{\url{https://satoshidice.com/}}.


\bibitem[wal(2022)]%
        {walletexplorer}
 \bibinfo{year}{2022}\natexlab{}.
\newblock \bibinfo{title}{Wallet Explorer}.
\newblock
\newblock
\newblock
\shownote{\url{https://www.walletexplorer.com/info}}.


\bibitem[was(2022)]%
        {wasabi}
 \bibinfo{year}{2022}\natexlab{}.
\newblock \bibinfo{title}{Wasabi Wallet}.
\newblock
\newblock
\newblock
\shownote{\url{https://wasabiwallet.io/}}.


\bibitem[wat(2022)]%
        {watchyourback}
 \bibinfo{year}{2022}\natexlab{}.
\newblock \bibinfo{title}{Watch your back}.
\newblock
\newblock
\newblock
\shownote{\url{https://github.com/cybersec-code/watchyourback}}.


\bibitem[Androulaki et~al\mbox{.}(2013)]%
        {evaluatingAndroulaki}
\bibfield{author}{\bibinfo{person}{Elli Androulaki},
  \bibinfo{person}{Ghassan~O. Karame}, \bibinfo{person}{Marc Roeschlin},
  \bibinfo{person}{Tobias Scherer}, {and} \bibinfo{person}{Srdjan Capkun}.}
  \bibinfo{year}{2013}\natexlab{}.
\newblock \showarticletitle{{Evaluating User Privacy in Bitcoin}}. In
  \bibinfo{booktitle}{\emph{Financial Cryptography and Data Security}}.
\newblock


\bibitem[{Bartoletti} et~al\mbox{.}(2018)]%
        {ponziBartoletti}
\bibfield{author}{\bibinfo{person}{M. {Bartoletti}}, \bibinfo{person}{B.
  {Pes}}, {and} \bibinfo{person}{S. {Serusi}}.}
  \bibinfo{year}{2018}\natexlab{}.
\newblock \showarticletitle{{Data Mining for Detecting Bitcoin Ponzi Schemes}}.
  In \bibinfo{booktitle}{\emph{Crypto Valley Conference on Blockchain
  Technology}}.
\newblock


\bibitem[Beckett(2021)]%
        {bitcoinfogAffidavit}
\bibfield{author}{\bibinfo{person}{Devon Beckett}.}
  \bibinfo{year}{2021}\natexlab{}.
\newblock \bibinfo{title}{{Bitcoin Fog Affidavit by Internal Revenue Service}}.
\newblock
\newblock
\newblock
\shownote{\url{https://storage.courtlistener.com/recap/gov.uscourts.dcd.230456/gov.uscourts.dcd.230456.1.1_1.pdf}}.


\bibitem[Caballero et~al\mbox{.}(2022)]%
        {goodfatr}
\bibfield{author}{\bibinfo{person}{Juan Caballero}, \bibinfo{person}{Gibran
  Gomez}, \bibinfo{person}{Srdjan Matic}, \bibinfo{person}{Gustavo Sánchez},
  \bibinfo{person}{Silvia Sebastián}, {and} \bibinfo{person}{Arturo
  Villacañas}.} \bibinfo{year}{2022}\natexlab{}.
\newblock \showarticletitle{{GoodFATR: A Platform for Automated Threat Report
  Collection and IOC Extraction}}.
\newblock \bibinfo{journal}{\emph{CoRR}}  \bibinfo{volume}{abs/2208.00042}
  (\bibinfo{date}{July} \bibinfo{year}{2022}).
\newblock
\urldef\tempurl%
\url{https://doi.org/10.48550/arXiv.2208.00042}
\showDOI{\tempurl}


\bibitem[{CERT Polska}(2015)]%
        {slaveReport}
\bibfield{author}{\bibinfo{person}{{CERT Polska}}.}
  \bibinfo{year}{2015}\natexlab{}.
\newblock \bibinfo{title}{Case Study of Malicious Actors: Going Postal}.
\newblock
\newblock
\newblock
\shownote{\url{https://maldroid.github.io/docs/The_Postal_Group.pdf}}.


\bibitem[Christin(2013)]%
        {travelingChristin}
\bibfield{author}{\bibinfo{person}{Nicolas Christin}.}
  \bibinfo{year}{2013}\natexlab{}.
\newblock \showarticletitle{{Traveling the Silk Road: A measurement analysis of
  a large anonymous online marketplace}}. In \bibinfo{booktitle}{\emph{The
  World Wide Web Conference}}.
\newblock


\bibitem[Clayton et~al\mbox{.}(2015)]%
        {clayton2015concentrating}
\bibfield{author}{\bibinfo{person}{Richard Clayton}, \bibinfo{person}{Tyler
  Moore}, {and} \bibinfo{person}{Nicolas Christin}.}
  \bibinfo{year}{2015}\natexlab{}.
\newblock \showarticletitle{{Concentrating Correctly on Cybercrime
  Concentration}}. In \bibinfo{booktitle}{\emph{Workshop on Economics of the
  Information Society}}.
\newblock


\bibitem[Commission(2020)]%
        {gamblingUkKyc}
\bibfield{author}{\bibinfo{person}{UK~Gambling Commission}.}
  \bibinfo{year}{2020}\natexlab{}.
\newblock \bibinfo{title}{{LCCP Condition 17.1.1 - Customer identity
  verification}}.
\newblock
\newblock
\newblock
\shownote{\url{https://www.gamblingcommission.gov.uk/licensees-and-businesses/lccp/condition/17-1-1-customer-identity-verification}}.


\bibitem[Conti et~al\mbox{.}(2018)]%
        {economicConti}
\bibfield{author}{\bibinfo{person}{Mauro Conti}, \bibinfo{person}{Ankit
  Gangwal}, {and} \bibinfo{person}{Sushmita Ruj}.}
  \bibinfo{year}{2018}\natexlab{}.
\newblock \showarticletitle{On the Economic Significance of Ransomware
  Campaigns: {A} Bitcoin Transactions Perspective}.
\newblock \bibinfo{journal}{\emph{Computers \& Security}}  \bibinfo{volume}{79}
  (\bibinfo{year}{2018}), \bibinfo{pages}{162--189}.
\newblock
\showISSN{0167-4048}
\urldef\tempurl%
\url{https://doi.org/10.1016/j.cose.2018.08.008}
\showDOI{\tempurl}


\bibitem[De(2021)]%
        {bitcoinfogTakedown}
\bibfield{author}{\bibinfo{person}{Nikhilesh De}.}
  \bibinfo{year}{2021}\natexlab{}.
\newblock \bibinfo{title}{{US Officials Arrest Alleged Operator of \$336M
  Bitcoin Mixing Service}}.
\newblock
\newblock
\newblock
\shownote{\url{https://www.coindesk.com/us-officials-arrest-alleged-operator-of-336m-bitcoin-mixing-service}}.


\bibitem[Eisenkraft and Olshtein(2019)]%
        {ponyCheckpoint}
\bibfield{author}{\bibinfo{person}{Kobi Eisenkraft} {and} \bibinfo{person}{Arie
  Olshtein}.} \bibinfo{year}{2019}\natexlab{}.
\newblock \bibinfo{title}{Pony's {C\&C} servers hidden inside the Bitcoin
  blockchain}.
\newblock
\newblock
\newblock
\shownote{\url{https://research.checkpoint.com/2019/ponys-cc-servers-hidden-inside-the-bitcoin-blockchain/}}.


\bibitem[Europol(2019)]%
        {bestmixerTakedown}
\bibfield{author}{\bibinfo{person}{Europol}.} \bibinfo{year}{2019}\natexlab{}.
\newblock \bibinfo{title}{Multi-million euro cryptocurrency laundering service
  Bestmixer.io taken down}.
\newblock
\newblock
\newblock
\shownote{\url{https://www.europol.europa.eu/newsroom/news/multi-million-euro-cryptocurrency-laundering-service-bestmixerio-taken-down}}.


\bibitem[Gill and Taylor(2004)]%
        {gill2004preventing}
\bibfield{author}{\bibinfo{person}{Martin Gill} {and} \bibinfo{person}{Geoff
  Taylor}.} \bibinfo{year}{2004}\natexlab{}.
\newblock \showarticletitle{{Preventing Money Laundering or Obstructing
  Business? Financial Companies' Perspectives on ‘Know Your Customer’
  Procedures}}.
\newblock \bibinfo{journal}{\emph{British Journal of Criminology}}
  \bibinfo{volume}{44}, \bibinfo{number}{4} (\bibinfo{year}{2004}),
  \bibinfo{pages}{582--594}.
\newblock


\bibitem[Goldfeder et~al\mbox{.}(2017)]%
        {cookieGoldfeder}
\bibfield{author}{\bibinfo{person}{Steven Goldfeder}, \bibinfo{person}{Harry~A.
  Kalodner}, \bibinfo{person}{Dillon Reisman}, {and} \bibinfo{person}{Arvind
  Narayanan}.} \bibinfo{year}{2017}\natexlab{}.
\newblock \showarticletitle{When the cookie meets the blockchain: Privacy risks
  of web payments via cryptocurrencies}.
\newblock \bibinfo{journal}{\emph{PoPETs}}  \bibinfo{volume}{2018}
  (\bibinfo{year}{2017}), \bibinfo{pages}{179--199}.
\newblock


\bibitem[Harlev et~al\mbox{.}(2018)]%
        {DBLP:conf/hicss/HarlevYLMV18}
\bibfield{author}{\bibinfo{person}{Mikkel~Alexander Harlev},
  \bibinfo{person}{Haohua~Sun Yin}, \bibinfo{person}{Klaus~Christian
  Langenheldt}, \bibinfo{person}{Raghava~Rao Mukkamala}, {and}
  \bibinfo{person}{Ravi Vatrapu}.} \bibinfo{year}{2018}\natexlab{}.
\newblock \showarticletitle{Breaking Bad: De-Anonymising Entity Types on the
  Bitcoin Blockchain Using Supervised Machine Learning}. In
  \bibinfo{booktitle}{\emph{51st Hawaii International Conference on System
  Sciences, {HICSS} 2018, Hilton Waikoloa Village, Hawaii, USA, January 3-6,
  2018}}, \bibfield{editor}{\bibinfo{person}{Tung Bui}} (Ed.).
  \bibinfo{publisher}{ScholarSpace / {AIS} Electronic Library (AISeL)},
  \bibinfo{pages}{1--10}.
\newblock
\urldef\tempurl%
\url{http://hdl.handle.net/10125/50331}
\showURL{%
\tempurl}


\bibitem[Harley and Matrosov(2011)]%
        {gluptebaEarly}
\bibfield{author}{\bibinfo{person}{David Harley} {and}
  \bibinfo{person}{Aleksandr Matrosov}.} \bibinfo{year}{2011}\natexlab{}.
\newblock \bibinfo{title}{TDL4 and Glupteba: Piggyback PiggyBugs}.
\newblock
\newblock
\newblock
\shownote{\url{https://www.welivesecurity.com/2011/03/02/tdl4-and-glubteba-piggyback-piggybugs/h}}.


\bibitem[Haslhofer et~al\mbox{.}(2016)]%
        {graphsenseHaslhofer}
\bibfield{author}{\bibinfo{person}{Bernhard Haslhofer}, \bibinfo{person}{Roman
  Karl}, {and} \bibinfo{person}{Erwin Filtz}.} \bibinfo{year}{2016}\natexlab{}.
\newblock \showarticletitle{{O Bitcoin Where Art Thou? Insight into Large-Scale
  Transaction Graphs}}. In \bibinfo{booktitle}{\emph{SEMANTiCS}}.
\newblock


\bibitem[Horejsi and Salgado(2019)]%
        {gluptebaTrendMicro}
\bibfield{author}{\bibinfo{person}{Jaromir Horejsi} {and}
  \bibinfo{person}{Joseph C~Chen Salgado}.} \bibinfo{year}{2019}\natexlab{}.
\newblock \bibinfo{title}{Glupteba Hits Routers and Updates C\&C Servers}.
\newblock
\newblock
\newblock
\shownote{\url{https://www.trendmicro.com/en_us/research/19/i/glupteba-campaign-hits-network-routers-and-updates-cc-servers-with-data-from-bitcoin-transactions.html}}.


\bibitem[{Huang} et~al\mbox{.}(2018)]%
        {trackingHuang}
\bibfield{author}{\bibinfo{person}{D.~Y. {Huang}}, \bibinfo{person}{M.~M.
  {Aliapoulios}}, \bibinfo{person}{V.~G. {Li}}, \bibinfo{person}{L.
  {Invernizzi}}, \bibinfo{person}{E. {Bursztein}}, \bibinfo{person}{K.
  {McRoberts}}, \bibinfo{person}{J. {Levin}}, \bibinfo{person}{K. {Levchenko}},
  \bibinfo{person}{A.~C. {Snoeren}}, {and} \bibinfo{person}{D. {McCoy}}.}
  \bibinfo{year}{2018}\natexlab{}.
\newblock \showarticletitle{{Tracking Ransomware End-to-end}}. In
  \bibinfo{booktitle}{\emph{IEEE Symposium on Security and Privacy}}.
\newblock
\urldef\tempurl%
\url{https://doi.org/10.1109/SP.2018.00047}
\showDOI{\tempurl}


\bibitem[Huntley and Nagy(2021)]%
        {gluptebaTakedown}
\bibfield{author}{\bibinfo{person}{Shane Huntley} {and} \bibinfo{person}{Luca
  Nagy}.} \bibinfo{year}{2021}\natexlab{}.
\newblock \bibinfo{title}{Disrupting the Glupteba operation}.
\newblock
\newblock
\newblock
\shownote{\url{https://blog.google/threat-analysis-group/disrupting-glupteba-operation/}}.


\bibitem[II et~al\mbox{.}(2019)]%
        {skidmapTrendMicro}
\bibfield{author}{\bibinfo{person}{Augusto~Remillano II},
  \bibinfo{person}{Jakub Urbanec}, {and} \bibinfo{person}{Wilbert~Luy Saias}.}
  \bibinfo{year}{2019}\natexlab{}.
\newblock \bibinfo{title}{{Skidmap Malware Uses Rootkit to Hide Mining
  Payload}}.
\newblock
\newblock
\newblock
\shownote{\url{https://www.trendmicro.com/en_us/research/19/i/skidmap-linux-malware-uses-rootkit-capabilities-to-hide-cryptocurrency-mining-payload.html}}.


\bibitem[Jourdan et~al\mbox{.}(2018)]%
        {Jourdan2018CharacterizingEI}
\bibfield{author}{\bibinfo{person}{Marc Jourdan}, \bibinfo{person}{Sebastien
  Blandin}, \bibinfo{person}{Laura Wynter}, {and} \bibinfo{person}{Pralhad
  Deshpande}.} \bibinfo{year}{2018}\natexlab{}.
\newblock \showarticletitle{{Characterizing Entities in the Bitcoin
  Blockchain}}.
\newblock \bibinfo{journal}{\emph{IEEE International Conference on Data Mining
  Workshops}} (\bibinfo{year}{2018}).
\newblock


\bibitem[Kalodner et~al\mbox{.}(2020)]%
        {blocksci}
\bibfield{author}{\bibinfo{person}{Harry Kalodner}, \bibinfo{person}{Malte
  Möser}, \bibinfo{person}{Kevin Lee}, \bibinfo{person}{Steven Goldfeder},
  \bibinfo{person}{Martin Plattner}, \bibinfo{person}{Alishah Chator}, {and}
  \bibinfo{person}{Arvind Narayanan}.} \bibinfo{year}{2020}\natexlab{}.
\newblock \showarticletitle{{BlockSci: Design and Applications of a Blockchain
  Analysis Platform}}. In \bibinfo{booktitle}{\emph{USENIX Security
  Symposium}}.
\newblock


\bibitem[Kimayong(2019)]%
        {masadReport}
\bibfield{author}{\bibinfo{person}{Paul Kimayong}.}
  \bibinfo{year}{2019}\natexlab{}.
\newblock \bibinfo{title}{Masad Stealer: Exfiltrating using Telegram}.
\newblock
\newblock
\newblock
\shownote{\url{https://blogs.juniper.net/en-us/threat-research/masad-stealer-exfiltrating-using-telegram}}.


\bibitem[Krebs(2022)]%
        {contiLeaks}
\bibfield{author}{\bibinfo{person}{Brian Krebs}.}
  \bibinfo{year}{2022}\natexlab{}.
\newblock \bibinfo{title}{{Conti Ransomware Group Diaries, Part I: Evasion}}.
\newblock
\newblock
\newblock
\shownote{\url{https://krebsonsecurity.com/2022/03/conti-ransomware-group-diaries-part-i-evasion/}}.


\bibitem[Lee et~al\mbox{.}(2019)]%
        {cybercriminalLee}
\bibfield{author}{\bibinfo{person}{Seunghyeon Lee}, \bibinfo{person}{Changhoon
  Yoon}, \bibinfo{person}{Heedo Kang}, \bibinfo{person}{Yeonkeun Kim},
  \bibinfo{person}{Yongdae Kim}, \bibinfo{person}{Dongsu Han},
  \bibinfo{person}{Sooel Son}, {and} \bibinfo{person}{Seungwon Shin}.}
  \bibinfo{year}{2019}\natexlab{}.
\newblock \showarticletitle{{Cybercriminal Minds: An Investigative Study of
  Cryptocurrency Abuses in the Dark Web}}. In \bibinfo{booktitle}{\emph{Network
  and Distributed Systems Security Symposium}}.
\newblock


\bibitem[{Liao} et~al\mbox{.}(2016)]%
        {behindLiao}
\bibfield{author}{\bibinfo{person}{K. {Liao}}, \bibinfo{person}{Z. {Zhao}},
  \bibinfo{person}{A. {Doupe}}, {and} \bibinfo{person}{G. {Ahn}}.}
  \bibinfo{year}{2016}\natexlab{}.
\newblock \showarticletitle{{Behind Closed Doors: Measurement and Analysis of
  CryptoLocker Ransoms in Bitcoin}}. In \bibinfo{booktitle}{\emph{APWG
  Symposium on Electronic Crime Research}}.
\newblock


\bibitem[Liao et~al\mbox{.}(2016)]%
        {liao2016acing}
\bibfield{author}{\bibinfo{person}{Xiaojing Liao}, \bibinfo{person}{Kan Yuan},
  \bibinfo{person}{XiaoFeng Wang}, \bibinfo{person}{Zhou Li},
  \bibinfo{person}{Luyi Xing}, {and} \bibinfo{person}{Raheem Beyah}.}
  \bibinfo{year}{2016}\natexlab{}.
\newblock \showarticletitle{{Acing the IOC Game: Toward Automatic Discovery and
  Analysis of Open-Source Cyber Threat Intelligence}}. In
  \bibinfo{booktitle}{\emph{ACM SIGSAC Conference on Computer and
  Communications Security}}.
\newblock


\bibitem[{Lin} et~al\mbox{.}(2019)]%
        {evaluationLin}
\bibfield{author}{\bibinfo{person}{Y. {Lin}}, \bibinfo{person}{P. {Wu}},
  \bibinfo{person}{C. {Hsu}}, \bibinfo{person}{I. {Tu}}, {and}
  \bibinfo{person}{S. {Liao}}.} \bibinfo{year}{2019}\natexlab{}.
\newblock \showarticletitle{An Evaluation of Bitcoin Address Classification
  based on Transaction History Summarization}. In
  \bibinfo{booktitle}{\emph{2019 IEEE International Conference on Blockchain
  and Cryptocurrency (ICBC)}}.
\newblock


\bibitem[Maxwell(2013)]%
        {coinjoinMaxwell}
\bibfield{author}{\bibinfo{person}{Gregory Maxwell}.}
  \bibinfo{year}{2013}\natexlab{}.
\newblock \bibinfo{title}{CoinJoin: Bitcoin privacy for the real world}.
\newblock
\newblock
\newblock
\shownote{\url{https://bitcointalk.org/index.php?topic=279249.0}}.


\bibitem[McCoy et~al\mbox{.}(2012)]%
        {pharmaleaks}
\bibfield{author}{\bibinfo{person}{Damon McCoy}, \bibinfo{person}{Andreas
  Pitsillidis}, \bibinfo{person}{Jordan Grant}, \bibinfo{person}{Nicholas
  Weaver}, \bibinfo{person}{Christian Kreibich}, \bibinfo{person}{Brian Krebs},
  \bibinfo{person}{Geoffrey Voelker}, \bibinfo{person}{Stefan Savage}, {and}
  \bibinfo{person}{Kirill Levchenko}.} \bibinfo{year}{2012}\natexlab{}.
\newblock \showarticletitle{{PharmaLeaks: Understanding the Business of Online
  Pharmaceutical Affiliate Programs}}. In \bibinfo{booktitle}{\emph{USENIX
  Security Symposium}}.
\newblock


\bibitem[Meiklejohn et~al\mbox{.}(2013)]%
        {fistfulMeiklejohn}
\bibfield{author}{\bibinfo{person}{Sarah Meiklejohn}, \bibinfo{person}{Marjori
  Pomarole}, \bibinfo{person}{Grant Jordan}, \bibinfo{person}{Kirill
  Levchenko}, \bibinfo{person}{Damon McCoy}, \bibinfo{person}{Geoffrey~M.
  Voelker}, {and} \bibinfo{person}{Stefan Savage}.}
  \bibinfo{year}{2013}\natexlab{}.
\newblock \showarticletitle{{A Fistful of Bitcoins: Characterizing Payments
  among Men with No Names}}. In \bibinfo{booktitle}{\emph{Internet Measurement
  Conference}}.
\newblock


\bibitem[{Möser} et~al\mbox{.}(2013)]%
        {inquiryMoser}
\bibfield{author}{\bibinfo{person}{M. {Möser}}, \bibinfo{person}{R. {Böhme}},
  {and} \bibinfo{person}{D. {Breuker}}.} \bibinfo{year}{2013}\natexlab{}.
\newblock \showarticletitle{{An Inquiry into Money Laundering Tools in the
  Bitcoin Ecosystem}}. In \bibinfo{booktitle}{\emph{APWG eCrime Researchers
  Summit}}.
\newblock


\bibitem[Nagy(2020)]%
        {gluptebaSophos}
\bibfield{author}{\bibinfo{person}{Luca Nagy}.}
  \bibinfo{year}{2020}\natexlab{}.
\newblock \bibinfo{title}{Glupteba: Hidden Malware Delivery in Plain Sight}.
\newblock
\newblock
\newblock
\shownote{\url{https://news.sophos.com/wp-content/uploads/2020/06/glupteba_final.pdf}}.


\bibitem[Nakamoto(2008)]%
        {bitcoin}
\bibfield{author}{\bibinfo{person}{Satoshi Nakamoto}.}
  \bibinfo{year}{2008}\natexlab{}.
\newblock \bibinfo{title}{Bitcoin: A Peer-to-Peer Electronic Cash System}.
  (\bibinfo{year}{2008}).
\newblock
\newblock
\shownote{https://bitcoin.org/bitcoin.pdf}.


\bibitem[Paquet-Clouston et~al\mbox{.}(2019a)]%
        {ransomwareClouston}
\bibfield{author}{\bibinfo{person}{Masarah Paquet-Clouston},
  \bibinfo{person}{Bernhard Haslhofer}, {and} \bibinfo{person}{Benoit Dupont}.}
  \bibinfo{year}{2019}\natexlab{a}.
\newblock \showarticletitle{{Ransomware Payments in the Bitcoin Ecosystem}}.
\newblock \bibinfo{journal}{\emph{Journal of Cybersecurity}}
  \bibinfo{volume}{5}, \bibinfo{number}{1} (\bibinfo{year}{2019}).
\newblock


\bibitem[Paquet-Clouston et~al\mbox{.}(2019b)]%
        {spamsPaquetClouston}
\bibfield{author}{\bibinfo{person}{Masarah Paquet-Clouston},
  \bibinfo{person}{Matteo Romiti}, \bibinfo{person}{Bernhard Haslhofer}, {and}
  \bibinfo{person}{Thomas Charvat}.} \bibinfo{year}{2019}\natexlab{b}.
\newblock \showarticletitle{{Spams Meet Cryptocurrencies: Sextortion in the
  Bitcoin Ecosystem}}. In \bibinfo{booktitle}{\emph{ACM Conference on Advances
  in Financial Technologies}}.
\newblock


\bibitem[{Pletinckx} et~al\mbox{.}(2018)]%
        {cerber}
\bibfield{author}{\bibinfo{person}{S. {Pletinckx}}, \bibinfo{person}{C.
  {Trap}}, {and} \bibinfo{person}{C. {Doerr}}.}
  \bibinfo{year}{2018}\natexlab{}.
\newblock \showarticletitle{{Malware Coordination using the Blockchain: An
  Analysis of the Cerber Ransomware}}. In \bibinfo{booktitle}{\emph{IEEE
  Conference on Communications and Network Security}}.
\newblock


\bibitem[Portnoff et~al\mbox{.}(2017)]%
        {backpagePortnoff}
\bibfield{author}{\bibinfo{person}{Rebecca~S. Portnoff},
  \bibinfo{person}{Danny~Yuxing Huang}, \bibinfo{person}{Periwinkle Doerfler},
  \bibinfo{person}{Sadia Afroz}, {and} \bibinfo{person}{Damon McCoy}.}
  \bibinfo{year}{2017}\natexlab{}.
\newblock \showarticletitle{{Backpage and Bitcoin: Uncovering Human
  Traffickers}}. In \bibinfo{booktitle}{\emph{ACM SIGKDD International
  Conference on Knowledge Discovery and Data Mining}}.
\newblock


\bibitem[Ron and Shamir(2013)]%
        {quantitativeRon}
\bibfield{author}{\bibinfo{person}{Dorit Ron} {and} \bibinfo{person}{Adi
  Shamir}.} \bibinfo{year}{2013}\natexlab{}.
\newblock \showarticletitle{{Quantitative Analysis of the Full Bitcoin
  Transaction Graph}}. In \bibinfo{booktitle}{\emph{Financial Cryptography and
  Data Security}}.
\newblock


\bibitem[Ron and Shamir(2014)]%
        {dreadRon}
\bibfield{author}{\bibinfo{person}{Dorit Ron} {and} \bibinfo{person}{Adi
  Shamir}.} \bibinfo{year}{2014}\natexlab{}.
\newblock \showarticletitle{{How Did Dread Pirate Roberts Acquire and Protect
  his Bitcoin Wealth?}}. In \bibinfo{booktitle}{\emph{Financial Cryptography
  and Data Security}}.
\newblock


\bibitem[Rubín(2019)]%
        {clipsa}
\bibfield{author}{\bibinfo{person}{Jan Rubín}.}
  \bibinfo{year}{2019}\natexlab{}.
\newblock \bibinfo{title}{Clipsa - Multipurpose password stealer}.
\newblock
\newblock
\newblock
\shownote{\url{https://decoded.avast.io/janrubin/clipsa-multipurpose-password-stealer/}}.


\bibitem[Saias(2021)]%
        {skidmapAkamai}
\bibfield{author}{\bibinfo{person}{Evyatar Saias}.}
  \bibinfo{year}{2021}\natexlab{}.
\newblock \bibinfo{title}{Bitcoins, blockchains, and botnets}.
\newblock
\newblock
\newblock
\shownote{\url{https://blogs.akamai.com/sitr/2021/02/bitcoins-blockchains-and-botnets.html}}.


\bibitem[Spagnuolo et~al\mbox{.}(2014)]%
        {bitiodineSpagnoulo}
\bibfield{author}{\bibinfo{person}{Michele Spagnuolo},
  \bibinfo{person}{Federico Maggi}, {and} \bibinfo{person}{Stefano Zanero}.}
  \bibinfo{year}{2014}\natexlab{}.
\newblock \showarticletitle{{BitIodine: Extracting Intelligence from the
  Bitcoin Network}}. In \bibinfo{booktitle}{\emph{Financial Cryptography and
  Data Security}}.
\newblock


\bibitem[Taniguchi et~al\mbox{.}(2021)]%
        {pony}
\bibfield{author}{\bibinfo{person}{Tsuyoshi Taniguchi}, \bibinfo{person}{Harm
  Griffioen}, {and} \bibinfo{person}{Christian Doerr}.}
  \bibinfo{year}{2021}\natexlab{}.
\newblock \showarticletitle{{Analysis and Takeover of the Bitcoin-Coordinated
  Pony Malware}}. In \bibinfo{booktitle}{\emph{ACM ASIA Conference on Computer
  and Communications Security}}.
\newblock


\bibitem[Tekiner et~al\mbox{.}(2021)]%
        {sokCryptojacking}
\bibfield{author}{\bibinfo{person}{Ege Tekiner}, \bibinfo{person}{Abbas Acar},
  \bibinfo{person}{A.~Selcuk Uluagac}, \bibinfo{person}{Engin Kirda}, {and}
  \bibinfo{person}{Ali~Aydin Sel{\c{c}}uk}.} \bibinfo{year}{2021}\natexlab{}.
\newblock \showarticletitle{{SoK: Cryptojacking Malware}}. In
  \bibinfo{booktitle}{\emph{IEEE European Symposium on Security and Privacy}}.
\newblock


\bibitem[Tepper and Schmidt(2021)]%
        {forbesExchanges2021}
\bibfield{author}{\bibinfo{person}{Taylor Tepper} {and} \bibinfo{person}{John
  Schmidt}.} \bibinfo{year}{2021}\natexlab{}.
\newblock \bibinfo{title}{Best Crypto Exchanges For 2021}.
\newblock
\newblock
\newblock
\shownote{\url{https://www.forbes.com/advisor/investing/best-crypto-exchanges/}}.


\bibitem[{US Department of Justice}(2017)]%
        {btceTakedown}
\bibfield{author}{\bibinfo{person}{{US Department of Justice}}.}
  \bibinfo{year}{2017}\natexlab{}.
\newblock \bibinfo{title}{Russian National And Bitcoin Exchange Charged In
  21-Count Indictment For Operating Alleged International Money Laundering
  Scheme And Allegedly Laundering Funds From Hack Of Mt. Gox}.
\newblock
\newblock
\newblock
\shownote{\url{https://www.justice.gov/usao-ndca/pr/russian-national-and-bitcoin-exchange-charged-21-count-indictment-operating-alleged}}.


\bibitem[Yin and Vatrapu(2017)]%
        {firstSun}
\bibfield{author}{\bibinfo{person}{Haohua~Sun Yin} {and} \bibinfo{person}{Ravi
  Vatrapu}.} \bibinfo{year}{2017}\natexlab{}.
\newblock \showarticletitle{{A First Estimation of the Proportion of
  Cybercriminal Entities in the Bitcoin Ecosystem using Supervised Machine
  Learning}}. In \bibinfo{booktitle}{\emph{IEEE International Conference on Big
  Data}}.
\newblock


\bibitem[Zhu and Dumitras(2018)]%
        {chainsmith}
\bibfield{author}{\bibinfo{person}{Ziyun Zhu} {and} \bibinfo{person}{Tudor
  Dumitras}.} \bibinfo{year}{2018}\natexlab{}.
\newblock \showarticletitle{{ChainSmith: Automatically Learning the Semantics
  of Malicious Campaigns by Mining Threat Intelligence Reports}}. In
  \bibinfo{booktitle}{\emph{IEEE European Symposium on Security and Privacy}}.
\newblock


\end{thebibliography}
}

\appendix

\section*{Appendix}

\begin{table}[h]
\centering
\scriptsize
\begin{tabular}{|l l l|}
\hline
\textbf{Operation} & \textbf{Short} & \textbf{Address} \\
\hline
emptystring & 3J98t1 & 3J98t1WpEZ73CNmQviecrnyiWrnqRhWNLy \\
binance & 1NDyJt & 1NDyJtNTjmwk5xPNhjgAMu4HDHigtobu1s \\
\hline
cerber & 14ru2h & 14ru2hbZyJzXADef285wxLVCLEUwnMtmzT \\
cerber & 17gd1m & 17gd1msp5FnMcEMF1MitTNSsYs7w7AQyCt \\
cerber & 1CpTCV & 1CpTCVckjajNKDd7PsApV3cAkunVd4Mcmt \\
cerber & 1DiRvu & 1DiRvuCvL7rYKuaP8NLEH3bji34wUEC5US \\
cerber & 1GcnsL & 1GcnsLs7C31uuroNmUHwwbB5xQeNvm63Ee \\
cerber & 1HTDy9 & 1HTDy9SkfhwaNCXFA8wFCvN53f3iGpm8kb \\
cerber & 1ML94w & 1ML94w1SCudkiFHaEwYqTmKGTkywxVBuZg \\
\hline
glupteba & 15y7ds & 15y7dskU5TqNHXRtu5wzBpXdY5mT4RZNC6 \\
glupteba & 1CgPCp & 1CgPCp3E9399ZFodMnTSSvaf5TpGiym2N1 \\
glupteba & 1CUhaT & 1CUhaTe3AiP9Tdr4B6wedoe9vNsymLiD97 \\
glupteba & 34Rqyw & 34RqywhujsHGVPNMedvGawFufFW9wWtbXC \\
glupteba & 3NhC1b & 3NhC1bvy1dVoJseK7V54VgLFLVNyFuNTVM \\
\hline
pony & 19hi8B & 19hi8BJ7HxKK45aLVdMbzE6oTSW5mGYC82 \\
pony & 1BkeGq & 1BkeGqpo8M5KNVYXW3obmQt1R58zXAqLBQ \\
pony & 1CeLgF & 1CeLgFDu917tgtunhJZ6BA2YdR559Boy9Y \\
pony & 1GUghN & 1GUghNgnSbcnoR1CJwDkmco5e7MKfGBn8G \\
pony & 1N9ALZ & 1N9ALZUgqYzFQGDXvMY5j1c7PGMMGYqUde \\
pony & 1NL8QT & 1NL8QTrs1KEhcpmBJU947nbcPtSZcpTqQg \\
pony & bc1q0n & bc1q0nw2g0dgm6shk0xazmaq2tzmwke7ypsz4upzps \\
pony & bc1qh9 & bc1qh96q46mw72shp2j39uq3z0wh0gezguvk9qq5js \\
pony & bc1qwl & bc1qwlmhl95l8fnck0s9xg7kse4sqc5mvj237gdy47 \\
pony & bc1qzs & bc1qzsdd9flsg4l4kq3awt23vkmtc30n7skavkfle3 \\
pony & 35KxqL & 35KxqLRwQPf5a17sPGDXDcQ8G9TKTxat7m \\
pony & 3Dmcs6 & 3Dmcs6kKViugRBdtZwhdHTG4SampnURTbQ \\
\hline
phorpiextldr & phorpiextldr & 1Bn4JYKoVgQpZ73doWVFSNZBbwKj3cpJNR \\
phorpiextldr & A & 1GWuCbanfYFJDr76ZbRCEjcRmRvqEqUA98 \\
phorpiextldr & B & 1Bs5Hc7CT2PCCKQJEyVgRVyznGgjwMDMtu \\
phorpiextldr & C & 1YRVEawzapE3N8CPpXg46JbYd72LfP7nB \\
phorpiextldr & D & 12EMaHiZG75ztkjUjuPZhQDcyW89qRJVuR \\
phorpiextldr & E & 1FJXSVbNmBu7z3RTNGixrvXN1N7pdpRZF3 \\
phorpiextldr & F & 1G1aLEVzBcPzqV7bbzfcb8tyt4u7oUEZqZ \\
phorpiextldr & G & 1K4Rb5FP3bcgubJzvCLHJ8vVDye1rQ3UNh \\
phorpiextldr & H & 1CSDpCjyVHsuTb6i7zZ8dr81iUGL5ff7vM \\
phorpiextldr & I & 14wmNiq9y9quQm69Y1bXZ7HFRB5mu1o72r \\
phorpiextldr & J & 3GVBvLxVzDqBYxSKDi7uenKGw8xBpwvDgy \\
phorpiextldr & K & 1QJoFaenGQ3bunZkMuN7i7c7GpzWXhmNqo \\
phorpiextldr & L & 1Jkhy6eH9TPn7w9WxYK59a8FxP4KtKGgL4 \\
phorpiextldr & M & 1Ny3nHNMkkxhLzas2ABfbdr4ZNYunSV2bE \\
phorpiextldr & N & 13e32uVdf6TSH47G67igN1sSkbrV6QATSY \\
phorpiextldr & O & 3EEiBhtMgMzrrCnZnVxjKQnMWEeduYiyi5 \\
phorpiextldr & P & 3G5RektH9rucFBzemQQy5R9eB6RPtYehcX \\
phorpiextldr & Q & 3PkdsSUfjdz6wDUAiFJDYB4bJTwmEZGq8y \\
phorpiextldr & R & 3MTng3jxPihohziAfVSUu53BpifcSNT2EN \\
phorpiextldr & S & 1KmaM9GZC1xTBiUndYSstW4YSrVLbgdsEj \\
phorpiextldr & T & 1MWNkiFb3sqtmRwdEhPSkRYh3zUwusXWGD \\
\hline
mrpr0gr4mmer & A & 1QDNWjJdBnNp8JNuQFhRWeQXL3fDb84cVS \\
mrpr0gr4mmer & B & 1ASppKbVqcX1NxQrxsjNnkWLMMtQDnmvrx \\
mekotio	     & C & 159cFxcSSpup2D4NSZiuBXgsGfgxWCHppv \\
mrpr0gr4mmer & D & 1LWs9tvEmRqpiVYNny4CY912xnudVKsqXG \\ mrpr0gr4mmer & E & 187g2Ppkb8F5suyqZwSExnZcDN9VZY7JhK \\ mrpr0gr4mmer & F & 1E6qZkzbGZHh9hWF4dQcTUdbmsYkvBYPrR \\
mrpr0gr4mmer & G & 1M7aXuhpnyQ6r1xgjicNWHjzdnVLEDsMHh \\
mrpr0gr4mmer & H & 1BRkLxWE888QTpV1eMVVHheQDdZkmkhSyT \\ mrpr0gr4mmer & I & 1HTZmKHrjyFAJdtrE7P3gdmYLHjwpZpcYG \\ mrpr0gr4mmer & J & 1BquFCDcLsYyL1SRykPaEgukh5EekwMGJ1 \\
mrpr0gr4mmer & K & 35oEfNFn3myZJKskhyEuaNcuGj6XDvRJKy \\ mrpr0gr4mmer & L & 3B2j29iZJ64oT6VNrqw8m4EpVv6HPq4xaU \\ mrpr0gr4mmer & M & 3LvUwiGHn5DCvpFmnQhVrE4tNhPkftf4gL \\ mrpr0gr4mmer & N & 1Ga6NDFba13YSUq3GMz4hDQgzjTixZTFvb \\
mrpr0gr4mmer & O & 1KRAgKpA1QjdACNQmtZUFCViEPB5T57Qsj \\
\hline
\end{tabular}
\caption{Mapping between short name and full address.}
\label{tab:mapping}
\end{table}

\begin{figure}
  \centering
  \includegraphics[width=\columnwidth]{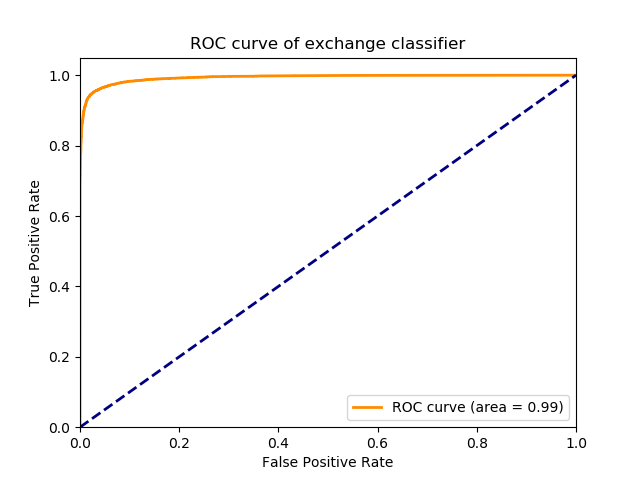}
  \caption{Exchange classifier ROC curve.}
  \label{fig:roc}
\end{figure}

\setlength{\tabcolsep}{3pt}
\begin{table}
\centering
\scriptsize
\begin{tabular}{|l r r r r r r|}
\hline
  \multicolumn{1}{|c}{} & \multicolumn{2}{c}{\textbf{Addresses}} & \multicolumn{2}{c}{\textbf{Funds Received}} & \multicolumn{2}{c|}{\textbf{Dates}} \\
  \textbf{Family} & \textbf{All} & \textbf{wTX} & \textbf{BTC} & \textbf{USD} & \textbf{Start} & \textbf{End} \\
\hline
	clipsa & 9,412 & 375 & 17.9741 & 277,514 & 2018-08-21 & 2021-02-23 \\
	cliptomaner & 1 & 1 & 16.1839 & 216,834 & 2019-11-17 & 2021-02-23 \\
	n40 & 1 & 1 & 27.4139 & 187,461 & 2017-05-31 & 2019-10-02 \\
	phorpiextldr & 15 & 15 & 19.2527 & 175,473 & 2018-05-31 & 2021-02-21 \\
	mrpr0gr4mmer & 3 & 3 & 7.3710 & 82,216 & 2020-01-04 & 2021-02-23 \\
	phorpiextrik & 8 & 7 & 11.4699 & 66,482 & 2016-08-14 & 2021-02-06 \\
	masad & 2 & 2 & 5.4072 & 46,850 & 2019-01-24 & 2021-02-20 \\
	clipbanker & 1 & 1 & 9.2829 & 36,425 & 2016-04-22 & 2021-02-20 \\
	cryptoshuffler & 1 & 1 & 23.5951 & 26,083 & 2016-09-27 & 2019-02-11 \\
	predatorthethief & 1 & 1 & 2.3695 & 22,539 & 2019-08-09 & 2020-11-15 \\
	azorult & 1 & 1 & 5.9413 & 21,872 & 2018-07-19 & 2019-03-24 \\
	mekotion40 & 3 & 3 & 1.4321 & 14,614 & 2018-01-29 & 2021-02-08 \\
	casbaneiro & 1 & 1 & 1.3246 & 8,922 & 2019-03-07 & 2021-01-07 \\
	protonbot & 2 & 2 & 0.9428 & 7,339 & 2018-04-17 & 2020-11-25 \\
	mispadu & 1 & 1 & 0.4752 & 5,235 & 2019-09-28 & 2021-01-12 \\
	mekotio & 3 & 2 & 0.3088 & 3,725 & 2020-06-25 & 2021-02-11 \\
	androidclipper & 1 & 1 & 0.1287 & 942 & 2018-02-23 & 2018-12-06 \\
	aggah & 1 & 1 & 0.0130 & 141 & 2020-05-18 & 2020-08-26 \\
	bitcoingrabber & 1 & 1 & 0.0082 & 71 & 2019-06-12 & 2019-10-28 \\
	\hline
	Total & 9,459 & 420 & 150.8949 & 1,200,738 & 2016-04-22 & 2021-02-23 \\
	\hline
	\hline
	locky & 7,076 & 7,074 & 16,353.7008 & 8,240,542 & 2016-01-14 & 2017-05-04 \\
	dmalocker & 10 & 10 & 1,131.7652 & 1,072,152 & 2015-12-28 & 2017-09-21 \\
	samsam & 45 & 20 & 435.6787 & 795,434 & 2016-01-13 & 2018-01-19 \\
	globeimposter & 3 & 1 & 594.4195 & 260,862 & 2014-11-23 & 2017-12-27 \\
	wannacry & 6 & 5 & 59.5967 & 118,460 & 2017-03-31 & 2021-02-05 \\
	cryptxxx & 1 & 1 & 94.8582 & 60,604 & 2016-06-04 & 2016-08-23 \\
		cryptotorlocker2015 & 1 & 1 & 5.6998 & 1,268 & 2015-01-18 & 2015-02-14 \\
	\hline
		Total & 7,142 & 7,112 & 18,675.7190 & 10,549,322 & 2014-11-23 & 2021-02-05 \\
\ignore{
n40 & 1 & 1 & 27.4138 & 187,461 & 2017-05-31 & 2019-10-02 \\
phorpiextldr & 15 & 15 & 19.2527 & 175,473 & 2018-05-31 & 2021-02-21 \\
clipsa & 9,412 & 375 & 17.9741 & 277,514 & 2018-08-21 & 2021-02-23 \\
cliptomaner & 1 & 1 & 16.1839 & 216,834 & 2019-11-17 & 2021-02-23 \\
phorpiextrik & 8 & 7 & 11.4698 & 66,482 & 2016-08-14 & 2021-02-06 \\
clipbanker & 1 & 1 & 9.2828 & 36,425 & 2016-04-22 & 2021-02-20 \\
azorult & 1 & 1 & 5.9412 & 21,872 & 2018-07-19 & 2019-03-24 \\
predatorthethief & 1 & 1 & 2.3695 & 22,538 & 2019-08-09 & 2020-11-15 \\
mekotion40 & 3 & 3 & 1.4321 & 14,614 & 2018-01-29 & 2021-02-08 \\
protonbot & 2 & 2 & 0.9427 & 7,339 & 2018-04-17 & 2020-11-25 \\
mispadu & 1 & 1 & 0.4752 & 5,235 & 2019-09-28 & 2021-01-12 \\
mekotio & 3 & 2 & 0.3088 & 3,725 & 2020-06-25 & 2021-02-11 \\
androidclipper & 1 & 1 & 0.1286 & 942 & 2018-02-23 & 2018-12-06 \\
aggah & 1 & 1 & 0.0130 & 141 & 2020-05-18 & 2020-08-26 \\
bitcoingrabber & 1 & 1 & 0.0082 & 71 & 2019-06-12 & 2019-10-28 \\
\hline
Total & 9,453 & 413 & 113.1970 & 1,036,668 & 2016-04-22 & 2021-02-23 \\
}
\hline
\end{tabular}
\caption{Clipper (top) and ransomware (bottom) profits measured on seeds.}
\label{tab:clippers}
\end{table}

\begin{table*}
\centering
\scriptsize
\begin{tabular}{|r|l|}
\hline
\textbf{Tags} &	\textbf{Source} \\
\hline

58606 & https://blog.talosintelligence.com/2018/10/anatomy-of-sextortion-scam.html \\
57014 & https://www.bitcoinabuse.com \\
37437 & https://github.com/MatteoRomiti/Deep\_Dive\_BTC\_Mining\_Pools/blob/master/dataset/miners.json \\
36548 & https://github.com/nopara73/WasabiVsSamourai/blob/master/WasabiVsSamourai/SamouraiCoinJoins.txt \\
10203 & https://www.cs.princeton.edu/~yuxingh/ransomware-public-data/cerber-locky-addresses.csv.txt \\
9412 & https://raw.githubusercontent.com/avast/ioc/master/Clipsa/appendix\_files/btc\_addresses\_complete.txt \\
7291 & https://zenodo.org/record/1238041 \\
6594 & https://bitcoin-otc.com/viewgpg.php \\
3907 & https://github.com/MatteoRomiti/Sextortion\_Spam\_Bitcoin \\
356 & https://www.walletexplorer.com \\
321 & https://www.justice.gov/opa/press-release/file/1304296/download \\
312 & https://raw.githubusercontent.com/VHRanger/tether/main/data/known\_addresses.json \\
295 & https://gist.github.com/MrChrisJ/4a959a51a0d2be356cc2e89566fc1d87 \\
223 & https://bitinfocharts.com \\
207 & https://bitcointalk.org/index.php?topic=576337 \\
121 & https://chain.info/richlist \\
115 & https://medium.com/meetbitfury/crystal-blockchain-analytics-investigation-of-the-zaif-exchange-hack-a3b4d1faed8f \\
115 & https://explorer.crystalblockchain.com \\
92 & https://wbtc.network/dashboard/audit \\
58 & https://research.checkpoint.com/2019/in-the-footsteps-of-a-sextortion-campaign \\

\hline
\end{tabular}
\caption{Top-20 tag source URLs.}
\label{tab:tags_source}
\end{table*}

\begin{table*}
\centering
\scriptsize
\begin{tabular}{| c| l | r r r | r r r r | r r | r | r|}
\cline{3-11}
	\multicolumn{2}{c|}{} & \multicolumn{3}{c|}{\textbf{Seeds}} & \multicolumn{4}{c|}{\textbf{Graph}} & \multicolumn{2}{c|}{\textbf{Classification}} & \multicolumn{2}{c}{} \\
\hline
	 \textbf{Class} & \textbf{Operation} 
	& \textbf{All} & \textbf{OW} & \textbf{Exp.}
	& \textbf{Comp.} & \textbf{Addr} & \textbf{Txes} & \textbf{Unexp.}
	& \textbf{Tagged}
	& \textbf{Exchanges} 
	& \textbf{Relations} 
  & \textbf{Runtime}\\
\hline
	\multirow{4}{*}{\rotatebox{90}{C\&C}}  
	&cerber		 &    4 &  0 &    4 &  1 &  448 &   904 &   0 &   5 &  320 & 0 & 1'00" \\
	&glupteba	 &    3 &  0 &    3 &  3 &   34 &    41 &   0 &   3 &    6 & 0 & 0'23" \\
	&pony		 &    4 &  0 &    4 &  2 &   27 & 1,548 &   1 &  17 &    2 & 0 & 0'33" \\
	&skidmap	 &    1 &  1 &    0 &  1 &    1 &     0 &   0 &   1 &    0 & 1 & - \\ \hline
	\multirow{4}{*}[-1.5cm]{\rotatebox{90}{Clipper}}
	&aggah		 &    1 &  0 &    1 &  1 &    7 &     8 &   1 &   3 &    2 & 1 & 0'21" \\
	&androidclipper	 &    1 &  1 &    0 &  1 &    1 &     0 &   0 &   1 &    0 & 1 & - \\ 	&azorult	 &    1 &  1 &    0 &  1 &    1 &     0 &   0 &   1 &    0 & 1 & - \\ 	&bitcoingrabber	 &    1 &  0 &    1 &  1 &   94 &   305 &   7 &  20 &    9 & 5 & 0'47" \\
	&casbaneiro	 &    1 &  0 &    1 &  1 &   59 &    96 &   2 &  32 &   11 & 4 & 1'03" \\
	&clipbanker	 &    1 &  1 &    0 &  1 &    1 &     0 &   0 &   1 &    0 & 1 & - \\ 	&clipsa  	 & 9412 &  0 &  375 & 33 &  681 &   840 &   2 & 469 &  124 & 6 & 1'22" \\
	&cliptomaner	 &    1 &  1 &    0 &  1 &    1 &     0 &   0 &   1 &    0 & 1 & - \\ 	&cryptoshuffler  &    1 &  0 &    1 &  1 &  507 & 2,331 &  18 & 134 &  180 &12 & 4'10" \\
	&masad		 &    2 &  0 &    2 &  1 &  471 & 1,478 &  22 &  99 &  170 &11 & 3'37" \\
	&mekotio	 &    3 &  0 &    2 &  2 &  146 &   385 &   8 &  35 &   53 &17 & 1'12" \\
	&mekotion40	 &    3 &  0 &    3 &  3 &  145 &   311 &   1 &  25 &   46 & 8 & 7'29" \\
	&mispadu	 &    1 &  1 &    0 &  1 &    1 &     0 &   0 &   1 &    0 & 1 & - \\ 	&mrpr0gr4mmer	 &    2 &  0 &    2 &  1 &  577 & 1,570 &   7 & 274 &  145 & 2 & 1'21" \\
	&n40		 &    1 &  1 &    0 &  1 &    1 &     0 &   0 &   1 &    0 & 1 & - \\ 	&phorpiex-tldr	 &   15 & 14 &    1 &  1 &   11 &    55 &   0 &   6 &    2 & 3 & 0'21" \\ 	&phorpiex-trik	 &    8 &  1 &    6 &  4 &  103 &   189 &   1 &  16 &   37 & 8 & 0'58" \\ 	&predatorthethief&    1 &  0 &    1 &  1 &  410 & 1,627 &  19 &  91 &  139 & 9 & 1'45" \\
	&protonbot	 &    2 &  2 &    0 &  1 &    2 &     0 &   0 &   2 &    0 & 2 & - \\ \hline
	\multirow{4}{*}[-0.25cm]{\rotatebox{90}{Ransomware}}
    &cryptotorlocker2015 &    1 &  0 &    1 &  1 &  147 & 2,137 &  17 &  60 &   23 & 8 & 115'22" \\
	&cryptxxx	 &    1 &  0 &    1 &  1 &  621 & 1,836 &   9 & 115 &  233 & 7 & 2'38" \\
	&dmalocker	 &   11 &  0 &   11 &  1 &  482 & 1,372 &   1 &  91 &  145 & 4 & 1'22" \\
	&globeimposter	 &    3 &  0 &    1 &  1 &  601 & 5,240 &  15 &  72 &  249 & 4 & 3'01" \\
	&locky		 & 7076 &  0 & 7074 &  1 & 7,987& 11,686&  29 &7,207&  200 &12 & 12'08" \\
	&samsam		 &   45 &  0 &   20 & 15 &  303 &   533 &   5 &  55 &   77 & 5 & 0'47" \\
	&wannacry	 &    6 &  0 &    5 &  2 &   30 &   520 &   0 &  13 &   10 & 0 & 0'24" \\
\hline
\end{tabular}
\caption{Summary of forward-only explorations for the \numoperations malware families.}
\label{tab:graphs-skipback}
\end{table*}

\begin{table*}
\centering
\scriptsize
\begin{tabular}{|l|r|lp{0.75\linewidth}|}
\hline
\textbf{Class} & \textbf{Operation} & \multicolumn{2}{l|}{\textbf{Relations Found}} \\
\hline
  \multirow{3}{*}{\rotatebox{90}{\makecell{C\&C}}}
  & cerber         & 0 & - \\
  & glupteba       & 0 & - \\
  & pony           & 0 & - \\
  & skidmap        & 1 & exchange:coincola* \\
\hline
  \multirow{17}{*}{\rotatebox{90}{Clipper}}
 &aggah            &  1 & exchange:binance \\
 &androidclipper   &  1 & exchange:luno.com* \\
 &azorult          &  1 & exchange:tradeogre* \\
 &bitcoingrabber   &  5 & exchange:bitmex, exchange:coinbase, exchange:okex, exchange:yobit.net, payment:payeer \\
 &casbaneiro	   &  4 & exchange:bleutrade.com, exchange:exmo, exchange:localbitcoins, payment:coinpayments \\
 &clipbanker       &  1 & exchange:coinbase* \\
 &clipsa           &  6 & exchange:binance, exchange:bitfinex, exchange:changenow, exchange:localbitcoins, mixer:wasabi-fee, payment:coinpayments \\
 &cliptomaner      &  1 & exchange:tradeogre* \\
 &cryptoshuffler   & 12 & exchange:bittrex, exchange:btc-e.com, exchange:btctrade.com, exchange:cryptopay.me, exchange:cubits.com, exchange:exmo, exchange:poloniex, exchange:spectrocoin, mining:btcc-pool, payment:bitpay.com, payment:coinpayments, payment:webmoney \\
 &masad            & 11 & exchange:any-cash, exchange:binance, exchange:bitzlato, exchange:coinbase, exchange:cryptonator, exchange:localbitcoins, exchange:whitebit, exchange:z-exchange.ru, gambling:1xbit, payment:payeer, payment:coinpayments \\
 &mekotio          & 17 & clipper:mekotion40*, exchange:binance, exchange:bitso, exchange:bittrex, exchange:btcturk, exchange:coinbase, exchange:hitbtc, exchange:localbitcoins, exchange:mercadobitcoin, exchange:paykrip, gambling:1xbit, miscabuse:mrpr0gr4mmer, mixer:wasabi-fee, payment:payeer, payment:coinpayments, service:gourl.io, service:muchbetter \\
 &mekotion40       & 8 & clipper:mekotio*, exchange:coingate, exchange:dsx-exchange, exchange:mercadobitcoin, gambling:foxbit, gambling:yolo-group, mixer:wasabi-fee, payment:coinpayments \\
 &mispadu          & 1 & exchange:mercadobitcoin* \\
 &mrpr0gr4mmer     & 2 & exchange:binance, payment:coinpayments \\
 &n40              & 1 & exchange:mercadobitcoin* \\
 &phorpiex-tldr    & 3 & exchange:cryptonator*, gambling:duckdice.io, misabuse:saveyourself \\
 &phorpiex-trik    & 8 & exchange:bitcoin.de, exchange:coinbase, exchange:cryptonator*, exchange:localbitcoins, exchange:matbea.com, exchange:poloniex, gambling:rollin.io, payment:webmoney \\
 &predatorthethief & 9 & clipper:masad, exchange:binance, exchange:bitzlato, exchange:btcbank.com.ua, exchange:cryptonator, exchange:localbitcoins, exchange:totalcoin, exchange:whitebit, payment:coinpayments \\
 &protonbot        & 2 & exchange:cryptonator*, exchange:hitbtc* \\
\hline
  \multirow{15}{*}{\rotatebox{90}{Ransomware}} 
&cryptotorlocker2015 & 8 & exchange:bitcoin.de, gambling:betcoin-dice, gambling:fortunejack.com, gambling:luckybit, gambling:satoshibones, gambling:satoshidice, mining:deepbit, tormarket:silkroad2market \\
 &cryptxxx         & 11 & exchange:bitcoin.de, exchange:bitfinex, exchange:bitflyer, exchange:btc-e.com, exchange:coinbase, exchange:coingate, exchange:coins.co.th, exchange:huobi, exchange:localbitcoins, exchange:okcoin.com, exchange:xapo \\
 &dmalocker        &  6 & exchange:btc-e.com, exchange:localbitcoins, exchange:okex, exchange:poloniex, gambling:cloudbet.com, payment:coinpayments \\  &globeimposter    &  7 & exchange:bittrex, exchange:btc-e.com, exchange:localbitcoins, gambling:coingaming.io, payment:bitpay.com, payment:easywallet, payment:webmoney \\
 &locky            & 15 & exchange:bitfinex, exchange:bitstamp, exchange:btc-e.com, exchange:coinbase, exchange:coins.co.th, exchange:exmo, exchange:localbitcoins, exchange:matbea.com, exchange:poloniex, exchange:xzzx.biz, gambling:cloudbet.com, mixer:bitcoinfog, payment:bitpay.com, payment:webmoney, tormarket:nucleusmarket \\
 &samsam           &  6 & exchange:bitcoin.de, exchange:cubits.com, exchange:localbitcoins, exchange:poloniex, exchange:yobit.net, tormarket:alphabaymarket \\
 &wannacry	   &  2 & exchange:changelly, exchange:shapeshift \\
\hline
\end{tabular}
\caption{Summary of relationships found with forward-only explorations.
Tags with an asterisk at the end were found by multi-input clustering on the seeds, the rest were found by the exploration.}
\label{tab:relations_skipback}
\end{table*}

\begin{algorithm}[t]
\footnotesize
\caption{Cerber Oracle}
\label{alg:cerber}
\begin{algorithmic}[1]
	\Procedure{oracle}{$addr$}
	\For{tx \textbf{in} addr.withdrawal\_txes}
		\State $ins\_outs \gets tx.ins\_count == tx.outs\_count == 1$
		\If{ins\_outs}
			\State $o \gets tx.outs[0].address$
			\State $n\_with \gets o.withdrawal\_txes.size$
			\State $n\_dep \gets o.deposit\_txes.size$
			\If{n\_with == n\_dep == 1}
				\State $tx\_r \gets o.withdrawal\_txes[0]$
				\State $txr\_ins \gets tx\_r.ins\_count$
				\State $txr\_outs \gets tx\_r.outs\_count$
				\If{txr\_ins == txr\_outs == 1}
					\State $r\_addr \gets tx\_r.outs[0].address$
					\If{r\_addr == addr}
						\State $r\_val \gets tx\_r.fee + tx\_r.out\_value$
						\If{tx.out\_value == r\_val}
							\State \textbf{return} $True$
						\EndIf
					\EndIf
				\EndIf
			\EndIf
		\EndIf
	\EndFor
	\State \textbf{return} $False$
	\EndProcedure
\end{algorithmic}
\end{algorithm}

\begin{algorithm}[t]
\footnotesize
\caption{Pony Oracle}
\label{alg:pony}
\begin{algorithmic}[1]
	\Procedure{oracle}{$addr$}
	\State $ips \gets []$
	\State $tx1, tx2 \gets None, None$
	\For{tx \textbf{in} addr.txes}
		\If{tx \textbf{in} addr.deposit\_txes}
			\If{tx.inputs.size <= 3 \textbf{and} tx.outputs.size <= 2}
				\State $value \gets tx.outs[addr].value$
				\If{tx1}
					\State $tx2, tx2\_value \gets tx, value$
					\State $ts1 \gets tx1.block.timestamp$
					\State $ts2 \gets tx2.block.timestamp$
					\State $delta \gets (ts2 - ts1).total\_seconds$
          \If{delta <= 3600}
            \State $ip \gets decode(tx1\_value, tx2\_value)$
            \If{ip \textbf{and} is\_public(ip)}
              \State ips.append(ip)
            \EndIf
            					\EndIf
					\State $tx1 \gets None$
				\Else
					\State $tx1, tx1\_value \gets tx, value$
				\EndIf
			\Else
				\State $tx1 \gets None$
			\EndIf
		\Else
			\State $tx1 \gets None$
		\EndIf
	\EndFor
		\State \textbf{return} $\frac{2*ips.size}{addr.deposit\_txes.size} \ge 0.5$
	\EndProcedure
\end{algorithmic}
\end{algorithm}

\begin{algorithm}[t]
\footnotesize
\caption{Glupteba Oracle}
\label{alg:glupteba}
\begin{algorithmic}[1]
	\Procedure{oracle}{$addr, keys$}
	\For{tx \textbf{in} addr.withdrawal\_txes}
		\For{o \textbf{in} tx.outs}
      \State $data \gets None$
			\If{hasattr(o.address, 'data')}
				\State $data \gets o.address.data.hex$
			\EndIf
			\If{data \textbf{and} len(data) >= 56}
				\For{key \textbf{in} keys}
					\State $iv \gets unhexlify(data[:24])$
					\State $tag \gets unhexlify(data[-32:])$
					\State $data \gets unhexlify(data[24:-32])$
					\State $plain\_text \gets aes\_gcm\_decrypt(iv, data, tag, key)$
					\If{plain\_text}
						\State \textbf{return} $True$
					\EndIf
				\EndFor
			\EndIf
		\EndFor
	\EndFor
	\State \textbf{return} $False$
	\EndProcedure
\end{algorithmic}
\end{algorithm}

\end{document}